\documentclass{aa}  

\usepackage[flushleft]{threeparttable}
\usepackage{float}
\usepackage{graphics}
\usepackage{graphicx}
\usepackage{amssymb, amsmath}
\usepackage{tikz}
\usetikzlibrary{graphs, positioning, quotes, shapes.geometric}
\usepackage{amstext}
\usepackage[T1]{fontenc}
\usepackage{mathrsfs}
\usepackage[normalem]{ulem}
\usepackage{natbib,twoopt}
\usepackage{hyperref} 
\usepackage{soul,xcolor}
\usepackage{booktabs}
\usepackage{rotating}
\usepackage{comment}
\usepackage{pdflscape}    % for landscape pages
\usepackage{algorithm}
\usepackage[noend]{algpseudocode}

\usepackage{txfonts}
\makeatletter
\def\@to{to}
\def\as     {\ifmmode {\rlap.}$\,$''$\,$\! \else ${\rlap.}$\,$''$\,$\!$\fi}
     \def\decsec  {\ifmmode {\rlap.}$\,$^{\rm s}$\,$\! \else ${\rlap.}$\,$^{\rm s}$\,$\!$\fi}\def\decss  {\ifmmode {\rlap.}$\,$^{\rm s}$\,$\! \else ${\rlap.}$\,$^{\rm s}$\,$\!$\fi}
\makeatother

\begin{document}

   \title{BASIL: Fast broadband line-rich spectral-cube fitting and image visualization via Bayesian quadrature}

   \author{Y. Lin
          \inst{1}
          \and
          M. Adachi
          \inst{2,4}
          \and
          S. Spezzano\inst{1}
          \and
          G. Edenhofer\inst{3}
          \and
          V. Eberle\inst{3}
          \and
          M.A. Osborne
          \inst{2}
          \and
          P. Caselli\inst{1}}

\institute{Max-Planck-Institut f{\"u}r Extraterrestrische Physik, Giessenbachstr. 1, D-85748 Garching bei M{\"u}nchen\\
              \email{ylin@mpe.mpg.de}
         \and
         Machine Learning Research Group, University of Oxford, Walton Well Road, Oxford, OX2 6ED, United Kingdom
        \and
        Max Planck Institute for Astrophysics, Karl-Schwarzschild-Strasse 1,
D-85748, Garching bei M{\"u}nchen
        \and
        Lattice Lab, Toyota Motor Corporation, 1200 Mishuku, Susono, Shizuoka, Japan\\
             }
\date{Received xx xx, 2024; accepted xx xx, 2024}

\abstract
{Mapping the spatial distributions and abundances of complex organic molecules in hot cores and hot corinos surrounding nascent stars is crucial for understanding the astrochemical pathways and the inheritance of prebiotic material by nascent planetary systems. However, the line-rich spectra from these sources pose significant challenges for robustly fitting molecular parameters due to severe line blending and unidentified lines.}
{We present an efficient framework, Bayesian Active Spectral-cube Inference and Learning (BASIL), for estimating molecular parameter maps -- excitation temperature, column density, centroid velocity, and line width -- for hundreds of molecules based on the local thermodynamic equilibrium (LTE) model, applied to wideband spectral datacubes of line-rich sources. The main aim is to allow the simultaneous fitting of hundreds of molecules to disentangle line blending issues and map the kinematic and abundance spatial distributions of the molecular parameter maps.}
{We adopted stochastic variational inference (SVI) to infer molecular parameters from spectra at individual positions, achieving a balance between fitting accuracy and computational speed. For obtaining parameter maps, instead of querying every location or pixel, we introduced an active learning framework based on Bayesian quadrature and its parallelization. Specifically, we assessed and selected the locations or pixels of spectrum that are most informative for estimating the entire set of parameter maps by training a Gaussian processes (GP) model. By greedily selecting locations with maximum information gain, we achieve sublinear convergence: the estimation error of the GP model for parameter maps drops rapidly in the early stages of iterations and then stabilizes. At this point, we can halt the fitting process, providing a fast and reasonably accurate visualization of the molecular parameter maps, while further accuracy is obtained through additional iterations of model training by querying more locations.}
{We benchmarked our algorithm using a synthetic spectral cube of 40,000 ($200 \times 200$) pixels, in which each pixel contains 138,016 frequency grids, and fit an LTE model with SVI to obtain four spectral parameters for a list of 117 molecules (117$\times$4 dimensions). Our algorithm is able to estimate 468 molecule parameter maps for 40,000 pixels in $\sim$180 hours (18 iterations of 50 parallel fittings, $\sim10$ hours per batch), achieving a comparable root mean square error across all data points. The full analysis is done on a high-memory server with multi-core CPUs. In contrast, a traditional MCMC fitting would take approximately $\sim 2\times 10^6$ hours to achieve the same level of accuracy, while requiring significant manual tuning.
In particular, with two iterations of $\sim$20 hours computational time, the GP model predicts parameter maps that are visually accurate. Additional training iterations provide progressively more accurate results. This quick visualization meets the demands of big data in modern astronomical surveys.}
{}


   \maketitle
%
%-------------------------------------------------------------------

\section{Introduction}

%hot core, hot corinos, importance of abundance distribution etc, 
The study of hot cores and hot corinos is pivotal for understanding the earliest stages of star formation and the remarkable chemical complexity that arises in these environments. Hot cores are dense, compact regions ($>$10$^{6}$ cm$^{-3}$) of warm gas ($>$100 K) and dust surrounding newly formed high-mass stars, while hot corinos are their low-mass counterparts around solar-type protostars (\citealt{Dishoeck98}; \citealt{Ceccarelli07}). These objects facilitate the sublimation of ice mantles from dust grains, driving a rich gas-phase chemistry that produces a wealth of complex organic molecules (COMs). The high temperatures and densities in hot cores and hot corinos further boost the formation of COMs through gas-phase reactions and grain-surface chemistry (\citealt{2009araa}; \citealt{Ceccarelli23}). The resulting molecular inventory includes prebiotically significant species such as glycolaldehyde and ethylene glycol (\citealt{Belloche13}; \citealt{Coutens15}; \citealt{Jorgensen16}).

Analyzing the abundances and isotopologue ratios of COMs in hot cores and hot corinos offers insights into the chemical pathways and physical conditions in these regions. The derivation of these quantities requires calculating the column density ($N_{\mathrm{mol}}$) of multiple molecular species with several transitions, together with the excitation temperature ($T_{\mathrm{ex}}$). Deuterium fractionation, measured through elevated D/H ratios, can be an important indication of the formation history and the temperature of the parental cloud (\citealt{Rodgers96}; \citealt{Ceccarelli14}; \citealt{Belloche16}). Similarly, fractionations of $^{12}$C/$^{13}$C carry information on the UV-radiation field (\citealt{Milam05}; \citealt{Jorgensen20}). As such, the spatial distribution of molecules is influenced by physical and chemical gradients. Together with column density and excitation temperature, the properties of centroid velocity ($V_{\mathrm{c}}$) and velocity linewidth ($\sigma_{V}$) of a certain molecular species may be imprinted by distinct underlying gas kinematics: the velocity patterns can exhibit features such as rotating disks (\citealt{Loomis18b}; \citealt{Booth21}), outflows and jets (\citealt{Lefloch17}; \citealt{Busch24}), as well as infalling gas flows or streamers (\citealt{Pineda12, Pineda23}; \citealt{Bianchi23}). Extracting full sets of parameters as mapping results is essential for understanding the formation mechanisms in detail, by dissecting their different origins and, eventually, tracing the inheritance of prebiotic material in nascent planetary systems (\citealt{Bianchi19}; \citealt{Drozdovskaya19}; \citealt{Chahine22}; \citealt{Codella22}).

The detection of new, complex molecules in hot cores and hot corinos continually challenges our understanding of interstellar chemistry, unveiling hitherto unknown formation pathways and expanding the boundaries of molecular complexity attainable in star-forming regions (e.g., \citealt{Belloche14}; \citealt{McGuire18}; and \citealt{Rivilla22}). These discoveries have profound implications for astrochemical models and our comprehension of the prebiotic chemistry inherited by nascent planetary systems (\citealt{2009araa}; \citealt{CC12}; \citealt{Jorgensen20}). From a theoretical perspective, the detection of new molecules and precise measurements of isotopic ratios provide critical constraints for astrochemical models, which aim to describe the complex interplay between gas-phase reactions, grain-surface chemistry, and physical processes such as desorption and shock-induced chemistry (\citealt{Garrod08}; \citealt{Vasyunin13}; \citealt{JS20}). Consequently, hot cores and hot corinos serve as indispensable templates for elucidating the chemistry in other warm, dense environments, such as the cores of giant molecular clouds, the regions immersed in shocks, and the circumnuclear regions of galaxies.

The wealth of complex molecules and high densities in hot cores and hot corinos leads to line-rich spectra with severe blending of transitions from numerous species across wide frequency ranges. This blending complicates fitting molecular parameters and abundances from observational data. Robust spectral line modeling is essential to disentangle overlapping contributions and accurately constrain quantities such as column densities and excitation temperatures \citep{2009araa, Manigand20, Coutens22}. Effective approaches require the simultaneous fitting of significant molecular species to avoid errors from neglecting blending or omitting relevant transitions. Sensitive datasets also facilitate identifying weakly emitting molecules, which are particularly prone to blending issues. Commonly, only relatively isolated transitions are used for detection; however, accurate characterization of major species is also necessary to identify these lines. Advanced line analysis techniques, coupled with high spectral resolution and modern imaging, are critical for obtaining precise molecular parameters from these complex spectra.

Wideband, sensitive submillimeter ALMA observations toward hot cores and hot corinos are routinely carried out. For extracting molecular parameter information from these observations, it is common practice to simplify the radiative transfer model by assuming local thermodynamic equilibrium (LTE) (more details in Sect. \ref{sec:ltemodel}) and to describe the line profile using Gaussian profiles, which result from thermal and turbulent broadening. Under LTE, the populations of molecular energy levels are described by the Boltzmann distribution, which allows us to express the molecular excitation solely as a function of temperature using
\begin{equation}
\frac{n_u}{n_l} = \frac{g_u}{g_l} e^{-\frac{E_{ul}}{kT}},
\end{equation}
where $n_{\mathrm{u}}$ and $n_{\mathrm{l}}$ are the populations of the upper and lower energy levels, $g_{\mathrm{u}}$ and $g_{\mathrm{l}}$ are their degeneracies, $E_{\mathrm{ul}}$ is the energy difference, and $T_{\mathrm{ex}}$ is the excitation temperature for this transition. This assumption is valid for the high-density environments of hot cores and hot corinos, where collisional rates dominate over radiative processes approaching thermalization. In practice, we often assume that all transitions are thermalized, resulting in equal $T_{\mathrm{ex}}$ for all transitions. This is, strictly speaking, the constant excitation temperature condition (e.g., \citealt{Caselli02ctex}), where the single excitation temperature is equivalent to the gas kinetic temperature, $T_{\mathrm{kin}}$. The distribution of populations is characterized by a partition function, $Q_{T_{\mathrm{ex}}}$, to link to the total gas column density. 
In particular, the spatial variations of molecular abundances are of great interest for understanding the formation and destruction pathways of molecules in certain physical environments. For enhancing astrochemical studies, it also requires the analysis of a large sample of observations in a systematic and unbiased way to yield statistics. It is desirable to conduct fast and robust derivations of the molecular physical parameters from these big datacubes. 

In the broader context of line decomposition tasks on spectral datacubes, several existing methodologies have emerged to tackle the challenges of line blending and parameter fitting. Techniques such as GAUSSPY$+$ (\citealt{Lindner15}, \citealt{gausspy}), ROHSA (\citealt{rohsa}), and SCOUSEPy (\citealt{scousepy}) have proven effective in addressing specific challenges, including overlapping Gaussian profiles or preserving spatial coherence. Bayesian frameworks, exemplified by AMOEBA (\citealt{Hafner21}) and pyspecnest (\citealt{Sokolov20}, \citealt{Ginsburg22}), allow for robust multiline fitting and the integration of prior knowledge, enhancing parameter constraints. Taking advantage of parallelization and graphics processing units (GPUs) in a recent study, \citealt{Juvela24} introduce a fast-fitting algorithm to handle large sets of spectra. Tools specifically tailored for analyzing line-rich datacubes of hot corinos -- with access to line catalogs for checking line blending, simultaneous fitting of multiple molecular species, and various fitting algorithms -- include software such as MADCUBA (\citealt{Martin19}) and MAGIX (\citealt{Mueller17}). While these tools have advanced the field significantly, the unique demands of hot corino surveys, with their exceptional spectral density and molecular diversity, necessitate further refinement and adaptation to derive accurate molecular parameters in these chemically rich environments. In this work, we establish an efficient and generalized workflow for obtaining molecular parameter maps for broadband line-rich datacubes. 

\section{Model spectrum calculation under LTE}\label{sec:ltemodel}

The one-component LTE model for a molecule has the following functional form:
\begin{equation}
    T_{\mathrm{mb}} = S (1-e^{-\tau_{\mathrm{total}}}) - T_{\mathrm{bg}} (\nu/\nu_{\mathrm{ref}})^{T_{\mathrm{slope}}}(1-e^{-\tau_{\mathrm{total}}}), \label{eq:lte}
\end{equation}
where $T_{\mathrm{bg}}$ is the background temperature, modulated by a frequency-dependent temperature slope of $T_{\mathrm{slope}}$. $\nu$ is the frequency in consideration and $\nu_{\mathrm{ref}}$ denotes a fiducial value. The core calculation of this one-component radiative transfer equation falls to the calculation of $\tau_{\mathrm{total}}$, which includes optical depth contributed from both line and dust continuum, $\tau_{\mathrm{total}}$ = $\tau_{\mathrm{dust}}$ + $\tau_{\mathrm{line}}$. For demonstration purposes, and also since in most cases the dust continuum level (as a function of frequency) is already subtracted to produce the spectral datacube, we assume $T_{\mathrm{bg}}$ = 2.73 K, $T_{\mathrm{slope}}$ = 0 and $\tau_{\mathrm{dust}}$ = 0, to focus on optimizing the molecular line parameters. So in this context, $\tau_{\mathrm{total}}$ reduces to $\tau_{\mathrm{line}}$, with $S$ standing for source function, adhering to\footnote{When $\tau_{\mathrm{dust}}$ is taken into account, the corresponding source function $S$ should follow a weighted average form: $S$ = $\frac{\tau_{\mathrm{line}}J(T_{\mathrm{ex}})+\tau_{\mathrm{dust}}J(T_{\mathrm{dust}})}{\tau_{\mathrm{line}}+\tau_{\mathrm{dust}}}.$}
\begin{equation}
S = J = \frac{h \nu}{k_{\mathrm{B}}}\frac{1}{e^{\frac{h \nu}{k_{\mathrm{B}}T_{\mathrm{ex}}}}-1}.
\end{equation}

Moreover, we consider only overlapping line components for calculating additive optical depth coming from one molecule, i.e., trapping or damping of radiation contributed from another blending molecule in an adjacent frequency range to a particular line component (signified by a certain set of $A_{\mathrm{ul}}$, $E_{\mathrm{up}}$, $g_{\mathrm{u}}$, $\nu_{0}$) in consideration is omitted. 

The molecular line constants, including Einstein A coefficients, $A_{\mathrm{ul}}$, lower energy level, $E_{\mathrm{l}}$, upper energy level degeneracy, $g_{\mathrm{u}}$, and the rest frequency $\nu_{0}$ for all transitions of molecules of interest were first collected and saved to a look-up table, together with the partition functions, $Q_{T_{\mathrm{ex}}}$, at different excitation temperatures. The Cologne Database for Molecular Spectroscopy (CDMS, \citealt{Endres16}) and Lille Spectroscopic Database\footnote{\url{https://lsd.univ-lille.fr/}} (LSD) are used for collecting these constants. We turn to the JPL catalog\footnote{\url{https://spec.jpl.nasa.gov/home.html}} when a particular entry only exists there. 

The composite spectrum for a list of molecules of interest then follows,
\begin{equation}
T_{\mathrm{comp}} = \sum_{i} T_{\mathrm{mb, mol_{\mathrm{i}}}}, \label{eq:compsp}
\end{equation}
where mol$_{i}$ represents the $i$th molecule in the list. The $\tau_{\mathrm{line}}$ follows,
\begin{equation}
\tau_{\mathrm{line, mol_{\mathrm{i}}}} = \sum_{k} \frac{c^{2}N_{\mathrm{mol}}}{8 \pi \nu^{2} Q_{{T_{\mathrm{ex}}}}} A_{\mathrm{ul}} g_{u} e^{-E_{l}/T_{\mathrm{ex}}} (1-e^{-\frac{h \nu}{k_{B}T_{\mathrm{ex}}}})\phi(\nu_{0, \mathrm{shift}}, \sigma_{\nu_{0, \mathrm{shift}}}), \label{eq:tauline}
\end{equation}
where $\phi (\nu_{0}, \sigma_{\nu_{0}})$ denotes a Gaussian profile centered at rest frequency, $\nu_{0}$, shifted to $\nu_{0, \mathrm{shift}}$ = $\nu_{0}$ (1-$v_{c}$/$c$), and with a standard deviation, $\sigma_{\nu_{0, \mathrm{shift}}}$, characterized by $\sigma_{v}$, $\sigma_{\nu_{0, \mathrm{shift}}}$ = $\sigma_{v}$ $\nu_{0, \mathrm{shift}}$/c. $k$ stands for the $k$th transition with a constant set of $A_{\mathrm{ul}}$, $E_{\mathrm{up}}$, $g_{\mathrm{u}}$, and $\nu_{0}$, and spans all the transitions from molecule mol$_{i}$ in the frequency range in consideration. To summarize, the one-component (one-layer) LTE model for a single molecule is defined by four physical parameters: excitation temperature $T_{\mathrm{ex}}$, molecular column density $N_{\mathrm{mol}}$, centroid velocity $V_{\mathrm{c}}$, and line width $\sigma_{\mathrm{v}}$. In some contexts, source size, $\theta_{\mathrm{ss}}$, could be an additional parameter of interest, which affects the beam filling factor, $f_{\mathrm{bf}}$ = $\frac{\theta_{\mathrm{ss}}^{2}}{\theta_{\mathrm{ss}}^{2}+\theta_{\mathrm{beam}}^{2}}$, where $\theta_{\mathrm{beam}}$ denotes the beam size. The beam filling factor then acts as a scaling factor to $T_{\mathrm{mb, mol_{\mathrm{i}}}}$.

%--------------------------------------------------------------------
\section{The problem of wideband spectroscopic fitting and a unified approach}\label{sec:prob_set}
Fitting wideband observational spectroscopic surveys can be regarded as a three-layer problem. Due to the abundantly existing molecules, especially the COMs exhibiting complex energy levels with numerous transitions, it is best practice to simultaneously fit all molecules of interest to automatically consider the contribution of adjacent line components in the blending. We hereafter refer to the synthetic model spectrum composed of line emission from a list of molecules as a composite spectrum, calculated from LTE models (Sect. \ref{sec:ltemodel}). Within a wide frequency coverage of several tens of GHz at high frequency ($\gtrsim$300\,GHz), COMs typically exhibit a few thousand transitions, making the calculation of the composite spectrum computationally intensive. The efficient calculation of composite spectra for, for example, several tens or hundreds of molecules (the majority of which are COMs), constitutes the first layer of the task.
%%%
For a single molecular species, the LTE model is characterized by four parameters: excitation temperature, $T_{\mathrm{ex}}$, molecular column density, $N_{\mathrm{mol}}$, and two velocity parameters: centroid velocity, $V_{\mathrm{c}}$, and line width, $\sigma_{\mathrm{v}}$. Therefore, simultaneously fitting even tens of molecules involves hundreds of parameter dimensions, while high-dimensional optimization is a conventionally difficult problem. Moreover, considering the wideband spectroscopic data with sufficient frequency/velocity resolution, the obtained spectrum typically has over 100 000 data points; the large dataset adds another layer of difficulty in the optimization. The computational cost of simple point-estimate methods in the quasi-Newton family, such as the limited-memory Broyden-Fletcher-Goldfarb-Shanno algorithm (\citealt{Zhu97}, hereafter L-BFGS, and its variant L-BFGS-B), can be more manageable than (nonlinear) least-square methods, but they are prone to local minima and the corresponding parameter uncertainties (standard deviation) are ill-defined. Optimizing hundreds of parameters for a single spectrum in an efficient and relatively robust manner is the second layer of the task.

The third layer derives spatially continuous parameter maps from the wideband datacube. Obtaining a continuous distribution of COMs requires robustly constraining parameters for typically tens of thousands to millions of pixels. Although this process can be easily distributed in modern clusters, treating each pixel independently tends to produce parameter maps that show great discontinuity, which is unphysical and usually a result of optimizations trapped in local minima for certain pixels. To resolve this, it is necessary to incorporate spatial correlation (or assume the parameter maps are spatially smooth) as prior knowledge when fitting individual pixels, or to adopt a method that inherently ensures spatial continuity.

To this end, we propose a framework with separable steps to address this three-layer problem. When fitting spectral cubes to obtain parameter maps, instead of treating each pixel individually, we can regard the distributions of best-fit parameters as 2D functions of spatial locations. This approach enables capturing spatial correlations, which represent both smooth, continuous distributions and more complex, potentially fractal-like structures. To model and optimize these spatial distribution functions effectively, we propose using Bayesian active learning (BAL, \citealt{settles2009active, adachi2024adaptive}) combined with Gaussian process (GP, \citealt{Rasmussen}) models to construct these 2D functions. This method essentially reduces pixel-by-pixel (brute-force) sampling to sampling the most informative spatial locations, using GP predictions to recover the functions in question.

For fitting parameters of an individual spectrum, we turn to the algorithm family of variational inference (VI, \citealt{Blei16, Zhang17}), instead of traditional point-estimate optimization methods or Markov Chain Monte Carlo (MCMC) techniques. While MCMC methods provide accurate posterior distributions, they often struggle with scalability in high-dimensional problems. In contrast, VI algorithms offer a more scalable approach, maintaining the advantages of incorporating prior knowledge (in a more flexible way than that point-estimate methods such as L-BFGS) and quantifying uncertainties in the fitted parameters, while remaining computationally efficient for large-scale problems. It is worth noting that recent research has explored hybrid methods combining the strengths of MCMC and VI (e.g., \citealt{Salimans14}, \citealt{Habib18}). While not employed in this manuscript, such hybrid approaches could potentially offer a balance between accuracy and scalability in future work.

We leverage the strengths of three Python libraries—JAX, NumPyro, and SOBER—to construct the modeling pipeline. Specifically, the LTE composite spectrum model calculations are performed with JAX \citep{frostig2018compiling}, a high-performance numerical computing library that employs built-in just-in-time (JIT) compilation via Accelerated Linear Algebra (XLA) and automatic differentiation. We employ NumPyro (\citealt{NumPyro1}; \citealt{NumPyro2}), a probabilistic programming library built on top of JAX, to perform VI, specifically, the stochastic variational inference (SVI) method, to obtain the best-fit model parameters for individual spectra. 
Once the model parameters for batches of locations are inferred, we construct a GP model to estimate parameter maps. To select the most informative batch of spectra to be fit in each iteration, we adopt Bayesian quadrature (BQ, \citealt{basq1, adachi2023bayesian}). To this end, the SOBER library (\citealt{adachi2023sober, adachi2024a}) offers a plug-and-play modular framework to construct GP based on GPyTorch \citep{gardner2018gpytorch}. We also employ a Kronecker multi-output GP \citep{maddox2021bayesian} to account for correlations between parameter maps.

This synergistic combination of the state-of-the-art libraries of JAX, NumPyro, and SOBER not only streamlines the modeling process but also ensures that each step is executed efficiently and effectively. By leveraging the strengths of these libraries, we tackle computationally intensive calculations, perform accurate inference, and construct powerful nonparametric models within a cohesive and scalable framework, dissecting and solving the three-layer problem. We elaborate on the three major parts of the workflow in the following subsections, starting with a mathematical description of the problem setting.

\section{The problem setting and algorithm}\label{ref:prob_set_math}

\subsection{Problem setting}
We define the verifiable metric to gauge the performance of our proposed method for the problem setting described in Sect. \ref{sec:prob_set}. This metric provides a principled way to compare inferred parameters against known ground truths, enabling quantitative assessment of inference accuracy and model fidelity in a controlled setting.

\paragraph{Dataset definition}
Let $\mathcal{X}$ be the "2D map" domain where we wish to estimate the spatial distribution of spectral parameters. We define the 2D map as the discrete domain $\mathcal{X} \subseteq \mathbb{N}^{H \times W}$, where $H$ and $W$ are the number of spectra for height (with astronomical coordinates, as in declination--Dec.) and width (right ascension--R.A.) directions, respectively. The indices $i,j$ denote the position of the spectrum in the domain $\mathcal{X}$. Furthermore, we define $x_{i,j} \in \mathcal{X}$ as a "pixel"—the variable represents the location $i,j$. At a pixel $x_{i,j}$, there is a corresponding spectrum data $\nu_{i,j}$. By fitting the LTE model to $\nu_{i,j}$ using SVI, we obtain the estimated parameters. We have $N$ candidate molecules, and each molecule has four spectral parameters. Thus, there are $M = 4N$ spectral parameters to infer in total. Let $\mathcal{Y} \subseteq \mathbb{R}^{M}$ be the spectral parameters, and $y_{i,j}^{(k)} \in \mathcal{Y}$ be the $k$-th parameter we estimated by fitting $\nu_{i,j}$ via SVI. $k$ is the index of spectral parameters. Let us further define the datasets: let $\mathcal{D}_\text{all} = \{ x_{i,j}, y_{i,j}^{(k)}\}_{i,j,k=1}^{H, W, M}$ be the full datasets, $\mathcal{D}_\text{train} = \{ x_{i,j}, y_{i,j}^{(k)}\}_{i,j,k}^{H^\prime, W^\prime, M^\prime}$ be the training dataset, and $\mathcal{D}_\text{remain} = \mathcal{D}_\text{all} \backslash \mathcal{D}_\text{train}$ be the remaining dataset, where $\backslash$ is the set subtraction that removes the duplicating elements from the set (i.e., $\mathcal{D}_\text{all} = \mathcal{D}_\text{train} \cup \mathcal{D}_\text{remain}$).

\paragraph{Model definition}
The multi-output GP predicts the parameters $y_{i,j}^{(k)}$ at each pixel $x_{i,j}$ over the whole domain $\mathcal{X} \times \mathcal{Y}$. The GP model's predictive ability depends on the training dataset $\mathcal{D}_\text{train}$, and we wish to minimize the number of training data points $\lvert \mathcal{D}_\text{train} \rvert$ to achieve the same prediction accuracy. If the GP prediction is accurate enough, further training datasets are not needed.

\paragraph{Evaluation metric}
Our goal is to train the multi-output GP to minimize the prediction error of $y_{i,j}^{k}$ across the domain $\mathcal{X} \times \mathcal{Y}$ with as small a number of training data $\lvert \mathcal{D}_\text{train} \rvert$ as possible. There are two origins of prediction errors: the SVI inference error $r_{\mathrm{SVI}}$ and the GP prediction error $r_{\mathrm{GP}}$.

The SVI inference squared error is defined as
\begin{equation}
r_{\mathrm{SVI}}(x_{i,j}, k) = \left( y^{(k), *}_{i,j} - \hat y^{(k)}_{i,j} \right)^2,\label{eq:svi_err}
\end{equation}
where $y^{(k), *}_{i,j}$ is the ground-truth parameter and $\hat y^{(k)}_{i,j}$ is the inferred parameter by SVI. Here, we use maximum a posteriori (MAP) estimation as $\hat y^{(k)}_{i,j}$. While synthetic datasets (described later) can define the ground truth, $y^{(k), *}_{i,j}$, this is unobtainable in real-world datasets. The GP prediction squared error is expressed as follows:
\begin{equation}
r_{\mathrm{GP}}(x_{i,j}, k) =  \left( \hat y^{(k)}_{i,j} - \hat{f}^{(k)}(x_{i,j}) \right)^2, \label{eq:gp_err}
\end{equation}
where $\hat{f}^{(k)}$ is the MAP prediction from the GP model.

Now, our goal can be evaluated through the following root-mean-squared error, $\text{RMSE}$:
\begin{equation}
\epsilon_\mathrm{RMSE} = \sqrt{ \frac{1}{HWM}\sum_{\mathrm{i,j,k}}^{H,W,M} \left[
r_{\mathrm{SVI}}(x_{i,j}, k) + r_{\mathrm{GP}}(x_{i,j}, k)) \right]
}.\label{eq:rmse}
\end{equation}
In practice, we use the following ``estimated" RMSE:
\begin{equation}
\tilde{\epsilon}_\mathrm{RMSE} = \sqrt{ \frac{1}{HWM}\sum_{\mathrm{i,j,k}}^{H,W,M} r_{\mathrm{GP}}(x_{i,j}, k)
}, \label{eq:rmse_est}
\end{equation}
since the ground-truth parameters $y^{(k), *}_{i,j}$ are inaccessible and $r_{\mathrm{SVI}}$ cannot be obtained for real-world observatory data. This approach assumes that the SVI estimation error follows a zero-mean Gaussian distribution,
$$
    \hat{y}_{i,j}^{(k)} = f(x_{i,j}, k) + \epsilon_{i,j,k},
$$
where $y_{i,j}^{(k), *} = f(x_{i,j}, k)$ is the ground-truth molecular parameter map function (we assume this is a sample from GP), and $\epsilon_{i,j,k} \sim \mathcal{N}(0, \sigma_k^2)$ is the sum of two independent normally distributed random variables arising from SVI estimation and GP estimation errors. The sum of normally distributed random variables also follows a normal distribution, with the mean equal to the sum of individual means and the variance equal to the sum of individual variances. Since the means of the errors are zero, Eq.~\eqref{eq:rmse_est} remains Gaussian but with an increased variance $\sigma_k^2 = \mathbb{V}_x[r_{\mathrm{SVI}}(x, k)] + \mathbb{V}_x[r_{\mathrm{RMSE}}(x, k)]$.

\paragraph{Uncertainty estimate}
Here, the GP model can provide its uncertainty on prediction as predictive variance. Such uncertainty can be decomposed into two parts: aleatoric and epistemic uncertainty. Aleatoric uncertainty is the fundamental limitation of machine learning approaches, which is essentially the same as the experimental noise, i.e., $\mathbb{V}_x[r_\text{RMSE}(x, k)]$. Epistemic uncertainty is the limitation of the estimate accuracy due to the small number of observations and their diversity. While epistemic uncertainty is often referred to as reducible uncertainty, which can be reduced when collecting more data, aleatoric uncertainty is irreducible. Intuitively, as we collect more data, we can asymptotically reduce epistemic uncertainty to zero, then the predictive uncertainty reduces to the pure experimental noise, namely aleatoric uncertainty. Thus, we can view the GP predictive uncertainty $\mathbb{V}_{\hat{f}^{(k)}}[\hat{f}^{(k)}(x_{i,j})]$ as a proxy for RMSE, i.e., 
\begin{align}
\epsilon_\text{var} := \sum_{k=1}^M \mathbb{V}_{\hat{f}^{(k)}}[\hat{f}^{(k)}(x_{i,j})],
\end{align}
and $\lim_{i,j,k \rightarrow H,W,M} \epsilon_\text{var} = \tilde{\epsilon}_\text{RMSE} \geq 0$ due to the aleatoric uncertainty.
Our goal is to train the GP to have prediction variance below the given threshold, i.e., $\epsilon_\text{var} \leq \epsilon_\mathrm{thr}$, where $\epsilon_\mathrm{thr}$ is the stopping criterion. We can evaluate the efficacy of our algorithm by estimating how many training datasets are necessary for the $\epsilon_\text{var}$ to be below $\epsilon_\mathrm{thr}$. 

\subsection{Algorithm}

\begin{algorithm}[H]
\caption{BASIL: Bayesian Active Spectral-cube Inference and Learning}
\label{alg:bsv}
\begin{algorithmic}[1]
\normalsize
\State {Input: batch size $n$, stopping criterion $\epsilon$}
\State Set initial dataset $\mathcal{D}_\mathrm{train}$.
\While{$\epsilon_\text{var} > \epsilon_\mathrm{thr}$} 
\Comment{Stopping criterion}
\label{alg_line:stopping_criterion}
\State train GP model $f_t \leftarrow \mathrm{MLE}(f_{t-1}, \mathcal{D}_\mathrm{train})$.
\label{alg_line:train_gp}
\State Select next batch locations 
$X^n_t \leftarrow \mathrm{BASQ}(f, n)$
\label{alg_line:basq}
\State Parallel fitting $Y^n_t \leftarrow \mathrm{SVI}(X^n_t)$. \label{alg_line:svi_fitting}
\State Update dataset $\mathcal{D}_\mathrm{train} = \mathcal{D}_\mathrm{train} \cup (X^n_t, Y^n_t)$.
\EndWhile
\State Visualize 2D maps via GP prediction $f_t(X_\mathrm{all})$.
\end{algorithmic}
\end{algorithm}

Algorithm~\ref{alg:bsv} summarizes the flow of our proposed algorithm. On line~\ref{alg_line:train_gp}, we train a GP model by maximizing the marginal likelihood, also known as Type-II maximum likelihood estimation (MLE), following standard practices for GP training (see \citet{Rasmussen} for details). On line~\ref{alg_line:basq}, we employ BASQ~\citep{basq1} for parallel (or batch) Bayesian active learning, selecting $n$ batch locations based on the GP model’s predictive uncertainty—choosing a set of $n$ points that maximally reduce this uncertainty. On line~\ref{alg_line:svi_fitting}, we fit $n$ spectra using SVI in parallel via batch computation. This batch-sequential active learning process continues until the stopping criterion on Line~\ref{alg_line:stopping_criterion} is met. Finally, we visualize the molecular map using GP predictions for $N$ molecules, each with $k=4$ spectral parameters. Here, we denote $X^n_t := (x_{i,j})_{i,j=1}^{n} \subseteq X_\mathrm{remain}$ as the batch locations, and $Y^n_t := (y_{i,j}^{(k)})_{i,j,k=1}^{n,n,M}$ as the batch fitting results. 

\begin{figure*}[htb]
\centering
\begin{tikzpicture}[
    box/.style = {rectangle, draw, rounded corners, minimum width=6cm, minimum height=1cm, text centered, text width=5.5cm, fill=none},
    box_wide/.style = {rectangle, draw, rounded corners, minimum width=7cm, minimum height=1cm, text centered, text width=6.5cm, fill=none},
    decision/.style = {diamond, draw, minimum width=3cm, minimum height=1cm, text centered, text width=3cm, fill=none},
    arrow/.style = {thick, ->, >=stealth},
    process/.style = {rectangle, draw, rounded corners, minimum width=7cm, minimum height=2cm, text centered, text width=6.5cm, fill=none, dashed}
]

\node[box] (start) at (0,0) {Start};
\node[box] (input) at (0,-2) {Input: batch size $n$, $n$ randomly selected locations $X^n_0$\\stopping criterion $\epsilon$};
\node[box] (svi0) at (0,-4) {Parallel SVI Fitting Process\\$Y^n_0 \leftarrow \mathrm{SVI}(X^n_0)$};
\node[box] (init) at (0,-6) {Set initial dataset $\mathcal{D}_\mathrm{train} = (X^n_0, Y^n_0)$};
\node[box] (train0) at (0,-8) {Initial training of GP model via MLE:\\$f_t \leftarrow \mathrm{MLE}(f_{t-1}, \mathcal{D}_\mathrm{train})$};
\node[box] (check) at (0,-10) {Is $\epsilon_\text{var} > \epsilon_\mathrm{thr}$?};
\node[box] (select) at (0,-12) {Select next batch locations:\\$X^n_t \leftarrow \mathrm{BASQ}(f, n)$};
\node[process] (svi) at (0,-14) {Parallel SVI Fitting Process\\$Y^n_t \leftarrow \mathrm{SVI}(X^n_t)$};
\node[box] (update) at (0,-16) {Update dataset:\\$\mathcal{D}_\mathrm{train} = \mathcal{D}_\mathrm{train} \cup (X^n_t, Y^n_t)$};
\node[box] (train) at (0,-18) {Refined training of GP model via MLE:\\$f_t \leftarrow \mathrm{MLE}(f_{t-1}, \mathcal{D}_\mathrm{train})$};
\node[box] (vis) at (7,-12) {Visualize 2D maps via GP prediction $f_t(X_\mathrm{all})$};
\node[box] (end) at (7,-14) {End};

% Arrows
\draw[arrow] (start) -- (input);
\draw[arrow] (input) -- (svi0);
\draw[arrow] (svi0) -- (init);
\draw[arrow] (init) -- (train0);
\draw[arrow] (train0) -- (check);
\draw[arrow] (check) -- node[right] {Yes} (select);
%\draw[arrow] (train) -- (select);
\draw[arrow] (select) -- (svi);
\draw[arrow] (svi) -- (update);
\draw[arrow] (update) -- (train);
\draw[arrow] (train) -| (-6,-10) -- (check);  % Adjusted arrow path
\draw[arrow] (check) -|(7, -10)-- node[left] {No} (vis);
\draw[arrow] (vis) -- (end);

\end{tikzpicture}
\caption{Flow chart of BASIL.}
\label{fig:bsv_flowchart}
\end{figure*}
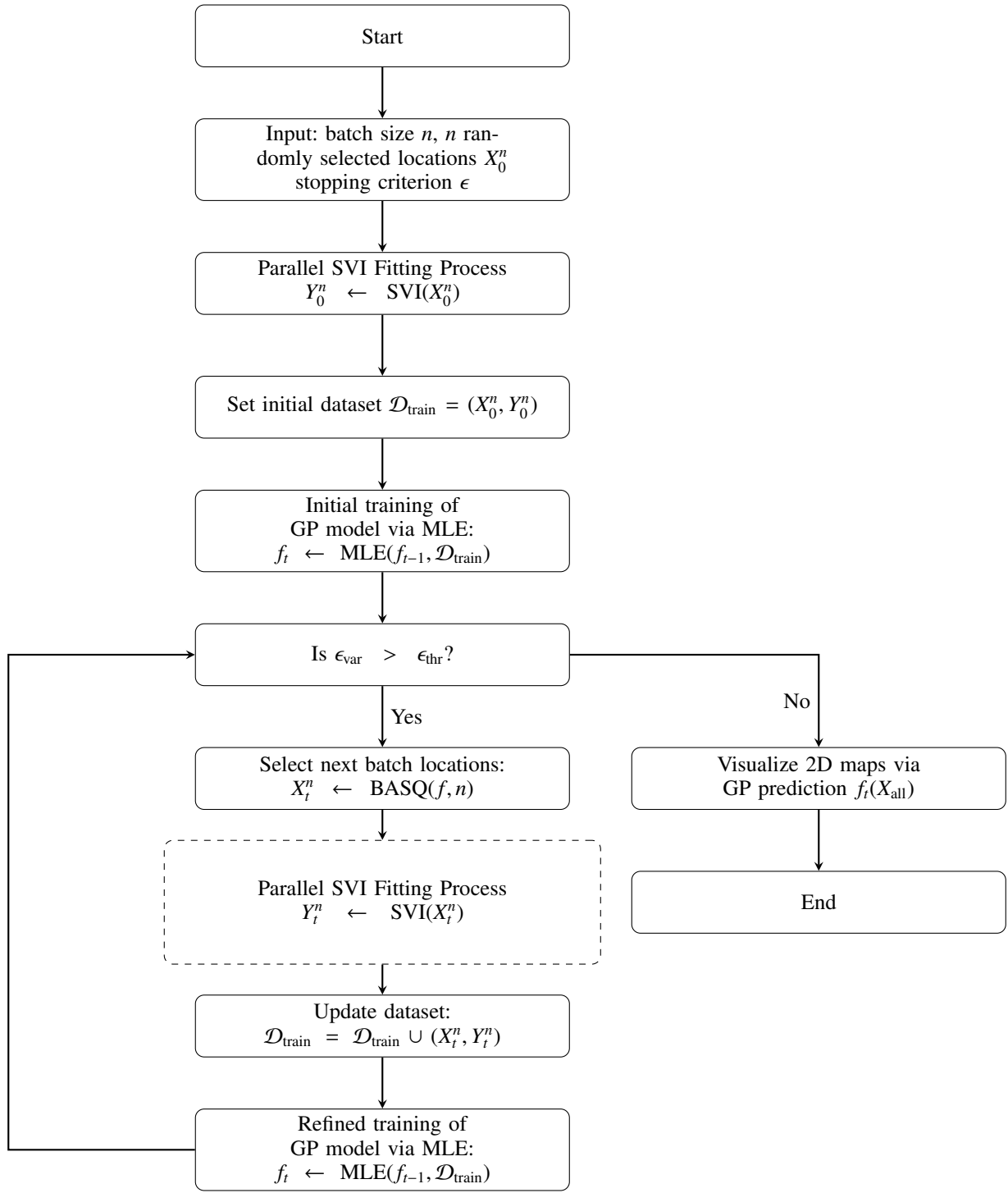

\section{Methods}
We are now ready to explain the details of each algorithmic component, the full workflow shown in Fig. \ref{fig:bsv_flowchart}, which is conducted on a CPU\footnote{The analysis was conducted on a server with four Intel Xeon Platinum 8260 CPUs (24 cores, 48 threads each, 2.40 GHz base clock) and 1.5 TiB of RAM. We observed that the SVI component ran more efficiently on a modern consumer CPU, likely due to higher single-thread performance, lower memory latency, and improved instruction set support utilized by JAX. Compared to the 10-hour runtime reported in the abstract based on the server configuration, this step can complete several times faster on newer hardware, depending on model complexity. }. In Sections~\ref{sec:jaxmodel} and \ref{sec:svi}, we outline the spectral fitting process described in Line~\ref{alg_line:svi_fitting}. Following this, we provide a detailed explanation of batch Bayesian active learning in Section~\ref{sec:gp}.

\subsection{Composite spectrum calculation with JAX}\label{sec:jaxmodel}

JAX\footnote{\url{https://jax.readthedocs.io/en/latest/}} is a Python library for high-performance numerical computing that employs JIT compilation to generate efficient machine code, enabling automatic differentiation for computing gradients. JAX seamlessly integrates with the Python ecosystem, making it a powerful tool that combines performance, differentiability, and Python's versatility for scientific computing and machine learning applications. It supports acceleration on GPUs and TPUs. We utilized several techniques in JAX to optimize the efficiency of calculating the LTE model for molecular line emission. For example, considering the form of the model calculation (Eq. \ref{eq:tauline}), we utilized dynamic slice updating for memory efficiency and vectorize the computation of $\tau_{\mathrm{line}}$ through the JAX's functions {\tt{vmap}} and {\tt{lax.scan}}.

\subsection{Deriving parameters for spectrum at a single location with stochastic variational inference}\label{sec:svi}

In order to infer molecular parameters for the composite spectrum from the observed spectrum, it is desirable to obtain a posterior density distribution that reflects the global probability of optimal solutions distributed over the parameter space. In probabilistic programming, Markov chain Monte Carlo sampling is a general method for obtaining posterior probability distributions; however, it is computationally heavy, especially when the dimension of the parameters is large (curse of dimensionality), and it also scales poorly with the size of the dataset. Although there are variants of the Hamiltonian Monte Carlo method that deal with subsamples of the dataset (\citealt{hmcecs}), the most efficient algorithms, such as the No-U-Turn Sampler (NUTS, \citealt{nuts}), perform worse as the dimension of the parameter space and data size grow larger, i.e., when the constraints of the parameter imposed by the data become poorer. In essence, MCMC uses the samples from a Markov chain to simulate the true posterior distribution. As an alternative inference method, VI attempts to approximate the posterior distribution (or conditional distribution) directly by probability distribution families that are easy to compute (\citealt{Wingate13}; \citealt{Blei16}). In this sense, instead of relying on sampling, VI is essentially an optimization task: the idea is to seek a simpler distribution or combination of distributions that most resembles the true posterior.

To characterize the similarity between the family of the approximate density distribution and the true posterior distribution, a criterion needs to be defined, and a common choice is the Kullback-Leibler (KL) divergence. In practice, instead of minimizing KL divergence, one commonly maximizes the evidence lower bound (ELBO) (essentially equivalent to the negative KL divergence) with an additional constant term, which represents a lower constraint on the logarithm of the marginal probability of the observations. Specifically, the formula for ELBO follows such a form of expectation as a function of our target parameters $y$ and latent parameters (\citealt{Ranganath14}),
\begin{equation*}
    \mathcal{L(\lambda)} = \mathbb{E}_{q(\lambda, z)}[\mathrm{log} p_{y}(x, z)-\mathrm{log} q(\lambda, z)],
\end{equation*}
where $x$ stands for observations, $p_{y}(x, z)$ the posterior distribution we want to approximate, and $\lambda$ are the variational parameters of a parameterized distribution $q(z)$, which is referred to as the variational distribution.

The complexity of the family of the variational distributions used to approximate the exact distribution determines the level of similarity that can be achieved. We use a multivariate normal variational distribution family, which are able to capture the correlations in the posterior, provided by the automatic guides in NumPyro. Choosing a distribution family comes with a trade-off between more accurately capturing the characteristics of the exact distribution and computational complexity. It is also possible to approximate a multimodal distribution using techniques such as bijective transformation to employ more expressive distribution families such as normalizing flows (\citealt{nice}; \citealt{JR15}).

The implementation of VI in NumPyro, specifically the SVI, combines the calculation of gradients and stochastic optimization (\citealt{Hoffman13}). The stochastic optimization means that in order to maximize the ELBO function, it draws a uniform random (Monte Carlo) sample from the variational distribution to get a noisy but unbiased estimate for the gradient of the ELBO (\citealt{Ranganath14}, \citealt{Wingate13}), iteratively updates local variational parameters, and then incorporates them into the update of the global variational parameters. This feature makes the method scale well with the size of the dataset.

\begin{figure}
    \centering
    \includegraphics[width=0.95\linewidth]{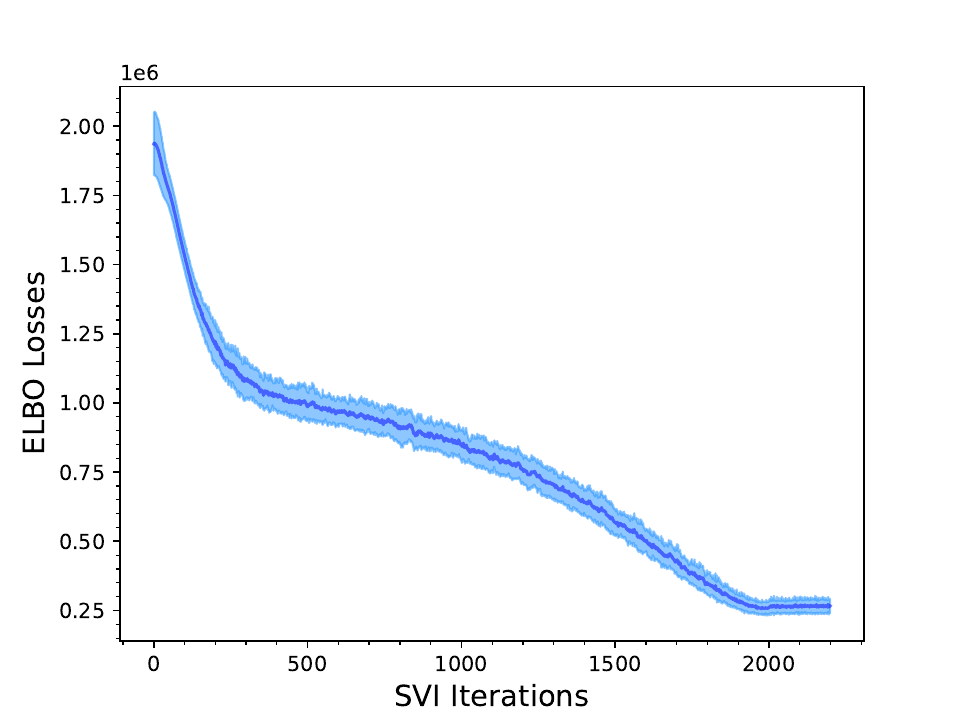}
    \caption{Examples of loss variation as a function of the number of iterations for 50 SVI batch fitting results, showing average losses and standard deviations (shaded area).}
    \label{fig:svi_losses}
\end{figure}

We adopted a chain of optimizers to find the optimal variational parameters enabled by library Optax\footnote{\url{https://github.com/google-deepmind/optax}} and a dynamic learning rate scheduler to achieve efficient convergence (\citealt{Loshchilov16}). Specifically, the learning rate starts from a large value and decays as the iterations proceed, such that in the beginning it tends to quickly reach the optimum location (especially in cases where there is not a reasonable initial position) and later on stabilizes. We iterated the ELBO optimization until it reached a plateau (see examples in Fig.\ref{fig:svi_losses}). It is also possible to apply informative initial positions of the model parameters to the SVI process; in practice, with a known list of molecules, a decent guess of the $N_{\mathrm{mol}}$ can be obtained, for example, by linear regression on the composite spectrum from a fixed set of $T_{\mathrm{ex, 0}}$, $N_{\mathrm{mol, 0}}$ for all molecules, and apply the resultant coefficients to $N_{\mathrm{mol, 0}}$ for individual molecules to obtain a scaled set of $N_{\mathrm{mol, 0}}$, to be used as a good initial position.

Unidentified lines compose an important factor complicating the line fitting procedures, which are either from unknown molecular species or from those not included in the molecule list of interest for the fitting. These unknown lines can be blended with the target lines and bias the optimization results. To take into account this effect, we introduced a likelihood of a mixture distribution to wrap around the model results (conditioned on the target parameters) and the observed spectrum, to categorize the emission channels as “inliers” and “outliers,” corresponding to a normal distribution and a heavy-tail Cauchy distribution, respectively. While the normal distribution for the “inliers” has a standard deviation largely determined by the observational noise, the heavy-tailed distribution is able to characterize channels that have line emission largely deviating from the model predictions (due to contribution from unidentified lines). We included the outlier probability as an additional parameter to optimize against. All the target parameters, i.e., sets of $N_{\mathrm{mol}}$, $T_{\mathrm{ex}}$, $v_{c}$, and $\sigma_{v}$ in respective ranges defined by lower and upper limits, are standardized (\citealt{RM15}).

\subsection{Sample-efficient spectral-cube parameter map approximation by Bayesian active learning}\label{sec:gp}
To achieve our goal, as described in Eq.\eqref{eq:rmse}, the easiest approach is to apply brute-force search -- querying every location in the dataset. This involves analyzing all spectra via LTE, as indicated by $\mathcal{D}_\text{train} = \mathcal{D}_\text{all}$. Consequently, the GP prediction error, $r_\text{GP}$, becomes zero at any pixel, resulting in the smallest possible RMSE. While this approach is the most accurate, it is also the least efficient in terms of computational time. Given that our spectral cube typically contains hundreds of thousands of pixels and numerous spectral cube datasets await analysis in modern line surveys, a balance between accuracy and computation time is necessary.
Ideally, we desire an algorithm that visualizes asymptotically accurate 2D parameter maps. Initially, the map should display a blurry yet globally accurate structure of the spatial physical parameters for each molecule, which is refined iteratively over time. The process stops when further iterations no longer result in significant changes to the image (user-defined stopping criteria). 

Bayesian active learning was invented to solve this exact problem. This algorithm selects the most informative pixel at each iteration, progressively making the visualization model more accurate. The simplest active learning method is known as the method of uncertainty sampling, which selects the pixel where the model is most uncertain in its prediction. This approach is theoretically well-known to approximate maximum information gain with high probability and has a sublinear convergence rate \citep{srinivas2009gaussian}. This means that prediction error drops rapidly in the beginning, and the rate of error decrease becomes smaller over time, eventually stabilizing in the later stages. Consequently, Bayesian active learning is a key enabler for achieving progressively accurate spectral-cube fitting for obtaining parameter maps.  For further efficiency, we introduced the following two techniques: (1) parallelization, and (2) incorporation of expert knowledge.

\paragraph{Parallelization}
A typical active learning scheme assumes a sequential setting, where the next fitting pixel is selected and its results are received one by one. In contrast, the parallel (or batch) setting allows for the selection of multiple pixels in a single iteration, running the fitting algorithm (SVI) in parallel, and receiving the results simultaneously. This approach is particularly useful when the fitting process is very time-consuming. SVI and, in general, all high-dimensional optimization problems, exemplify this case, taking several hours to fit one spectrum. For instance, if the fitting computation takes 2 hours per pixel, the sequential setting would require 20,000 hours to complete a parameter map of typically 10 000 pixels. 
However, with parallelization, assuming we can run 100 fitting computations in parallel, each two-hour iteration would fit 100 pixels, resulting in a total of 800 hours, or approximately 8 days. Thus, parallelization significantly accelerates the process.

While selecting a single informative point is straightforward—one simply chooses the global maximum of the GP predictive variance—parallel selection introduces additional complexity. For example, consider selecting two points. The first point would naturally coincide with the global maximum. However, the second point, chosen based on the next largest predictive variance, is likely to be close to the first due to the smoothness of the GP. This greedy selection leads to clustering near the initial maximum, which is suboptimal for reducing uncertainty.
This intuition suggests that the selected points should be spatially diverse. In statistics, such sampling strategies are often referred to as sparse sampling, diversified sampling, or repulsion sampling. In effect, parallel sampling requires solving an optimization problem that maximizes predictive variance while incorporating a "repulsion" constraint. A key question then arises: how should this constraint be optimally defined?

\citet{basq1} reformulated this challenging optimization problem as a quadrature problem. Conceptually, the goal is to select multiple points that effectively "cover" regions of high predictive variance across the domain. Formally, this task is referred to as quantization, which involves selecting the most representative points to best approximate the target distribution (\citealt{grayneuhoff}). Intuitively, to represent the target distribution, the selected points naturally spread out across the domain.
It is well-known that the quantization task is equivalent to solving the quadrature task. Thus, this parallelization task can be addressed using a quadrature solver; that is, parallelized uncertainty sampling, quantization, and quadrature are equivalent tasks. By reformulating the problem as a quadrature task, we can leverage existing algorithms that are both sample-efficient and computationally fast. For instance, \citet{basq1} employed the kernel quadrature algorithm introduced by \citet{hayakawa2022positively} (see their paper for detailed algorithmic descriptions).
More technically, Theorem 1 in \citet{basq1} demonstrated that this algorithm achieves an exponential convergence rate in the Gaussian case, which is a nearly rate-optimal result. As a result, this approach offers a provably efficient solution to the parallelization problem.

In our case, the (unnormalized) target distribution is the GP predictive uncertainty, $C_t(x, x)$, where $x \in \mathcal{D}_\text{remain}$. Note that this represents the sum of uncertainty across the entire parameter space; that is, $\sum{k=1}^M C_t^{(k)}(x, x)$. Fig.~\ref{fig:basq_sum_var} illustrates the sum of the GP predictive variance. Our algorithm generates randomized, sparse, and representative points over this distribution.

\paragraph{Incorporating expert knowledge}
As explained, the convergence rate depends on the spectral decay of the kernel Gram matrix, which reflects the GP model's predictive ability. To enhance GP prediction, we considered incorporating expert knowledge into the visualization process. Specifically, we leveraged the correlation between molecules and their physical parameters. Based on our current understanding of relations and formation pathways between different and/or chemically related molecules, one can intuitively judge the similarity of the physical parameters for the molecules in consideration, including the intrinsic correlation between $T_{\mathrm{ex}}$ and $N_{\mathrm{mol}}$ for individual molecules. Additionally, velocity parameter maps ($V_{\mathrm{c}}$ and $\sigma_{\mathrm{v}}$) for certain molecules, especially chemically related ones, often exhibit significant resemblance.
When high correlation is expected, we can set inter-parameter kernels, which can learn the correlation matrix from training data. Conversely, we can assume independence when correlation is not expected. Independence aids in reducing both memory usage and computational time. As the inter-parameter kernel computation can become prohibitive, particularly when the number of parameters is large (i.e., $M \gg 1$), assuming independence based on expert knowledge can significantly accelerate computation and save memory without sacrificing accuracy compared to a fully correlated model.
We adopted the Kronecker multi-output GP \citep{maddox2021bayesian} for the base model and created an interface allowing for specifications on which parameters might be correlated. This user-informed approach enables efficient and accurate spectral fitting while optimizing computational resources.
In a future work, an interactive knowledge elicitation framework (e.g., \citealt{adachi2024looping, xu2024principled}) can be more approachable for wider users.

This implementation offers a scalable and efficient approach for modeling correlated and independent tasks simultaneously within a GP framework, leveraging the Kronecker structure to address high-dimensional problems. It provides a flexible and principled way to capture the relationships between output dimensions, enabling more accurate and informative predictions in multitask learning scenarios.

\section{Test result with synthetic datacube}
\begin{figure*}
\hspace{-.5cm}\includegraphics[scale=0.3]{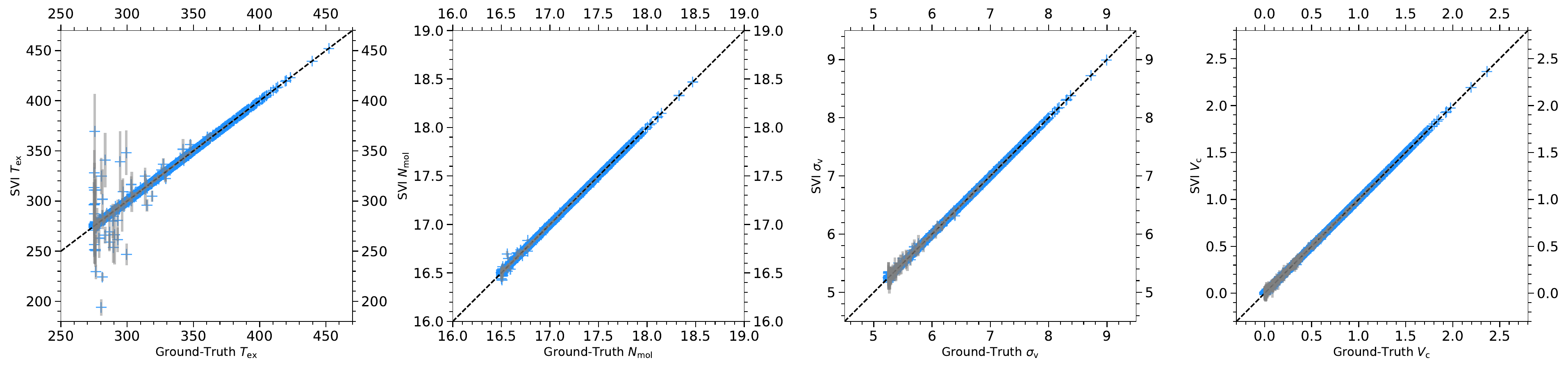}
\caption{Cross-validation results of $T_{\mathrm{ex}}$, $N_{\mathrm{mol}}$, $V_{\mathrm{c}}$, and $\sigma_{\mathrm{v}}$ for each molecule at 50 initial randomly selected positions.}
\label{fig:test_svi_92}
\end{figure*}

To demonstrate and fine-tune our proposed workflow, we used synthetic wideband spectral cubes for testing purposes. To allow for further benchmarking with observed data, for generating the synthetic data, we adopted the frequency range of the PILS survey (\citealt{Jorgensen16}) at ALMA band 6, covering 329.15-362.90 GHz, and a frequency resolution of 0.25 MHz, corresponding to 138 016 channels. We employed a list of 117 molecular entries (Table \ref{tab:mollist}) to generate synthetic data; these include simple molecules to COMs, including both main and rare isotopologues, and vibrational transitions for certain molecules are sometimes listed independently (consistent with the CDMS/JPL entries). We removed the molecules that have less than three transitions in the frequency range under consideration, which means that the SVI results for $N_{\mathrm{mol}}$ and $T_{\mathrm{ex}}$ cannot be properly constrained.

\begin{figure*}
    \includegraphics[width=\linewidth]{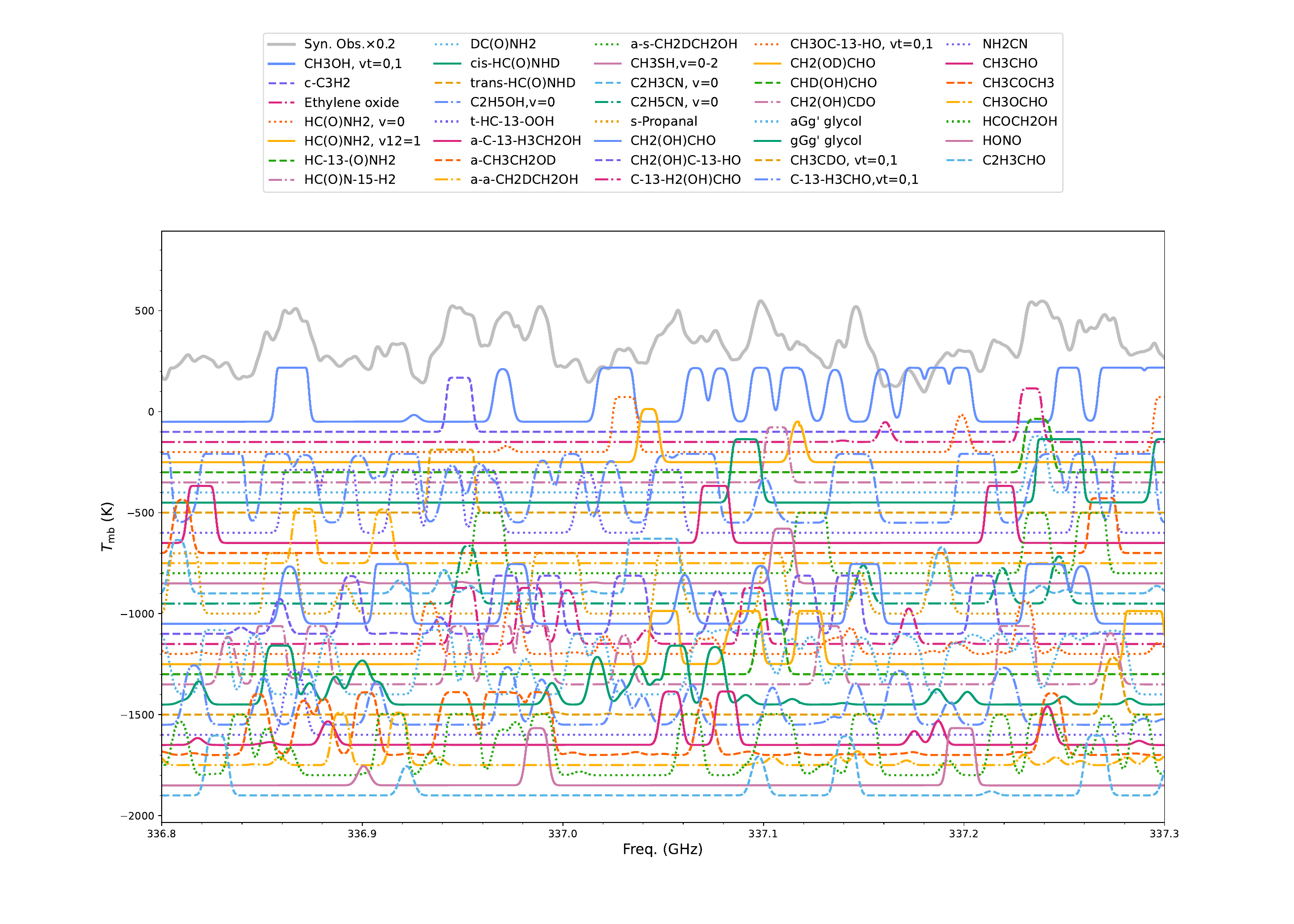}
    \label{fig:illu_line_bl}
    \caption{Example of a synthetic spectrum in the synthetic datacube, showing all molecules with significant emitting lines (peak intensity $>$250 K), as an illustration of line blending, offset by 50 K. The composite spectrum is scaled by 0.2 for better comparison.}
\end{figure*}

\begin{figure*}
    \includegraphics[width=\linewidth]{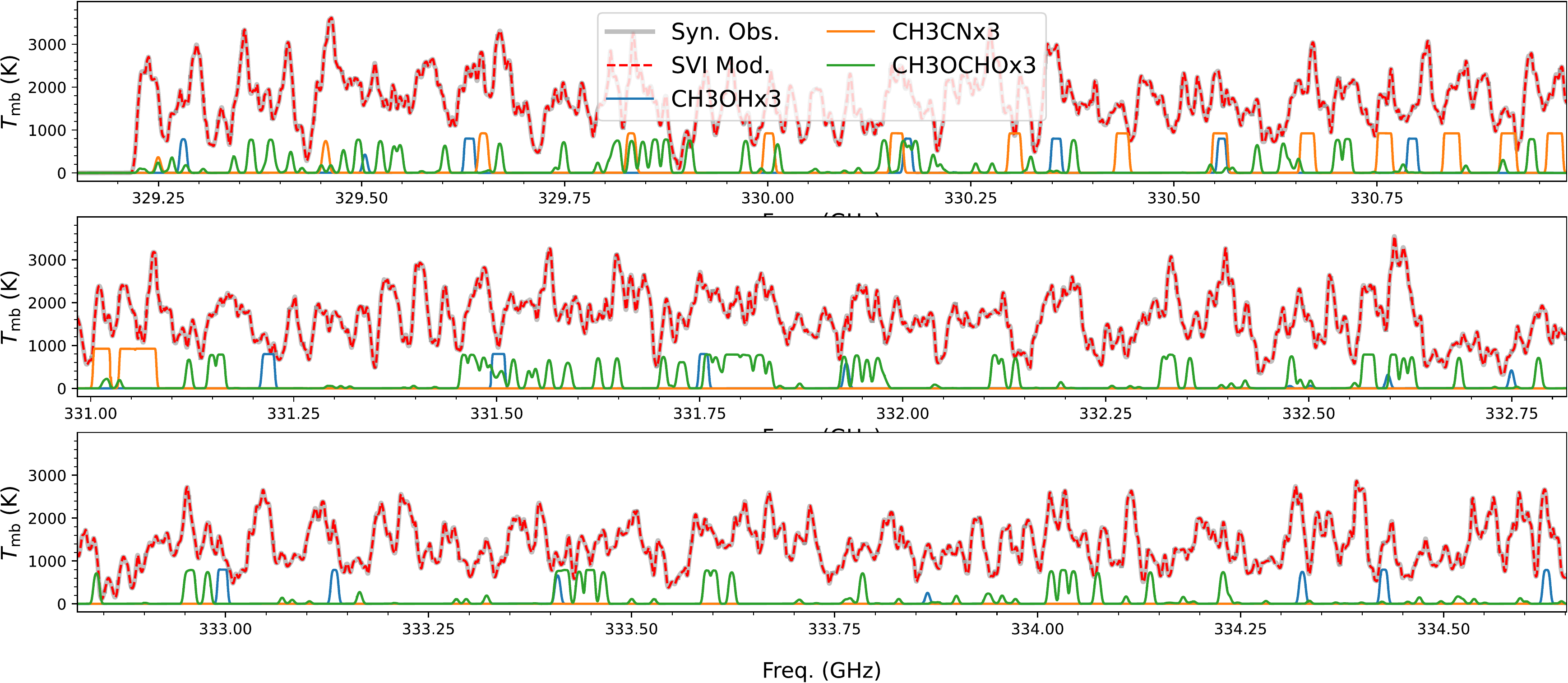}
    \caption{Example of a synthetic spectrum from the synthetic datacube. Subplots are arranged in chunks of increasing frequency. The synthetic observed spectrum is indicated as a thick gray line, and the SVI inferred model spectrum as a dashed red line. The synthetic spectra for the three molecules CH$_{3}$OH, CH$_{3}$CN, and CH$_{3}$OCHO, which compose the full synthetic observed spectrum along with the other 114 molecules, are scaled by a factor of three for illustration.}
        \label{fig:repre_sp1}

\end{figure*}

\subsection{Setup of the synthetic datacube}

\begin{figure*}
 \includegraphics[scale=0.65]{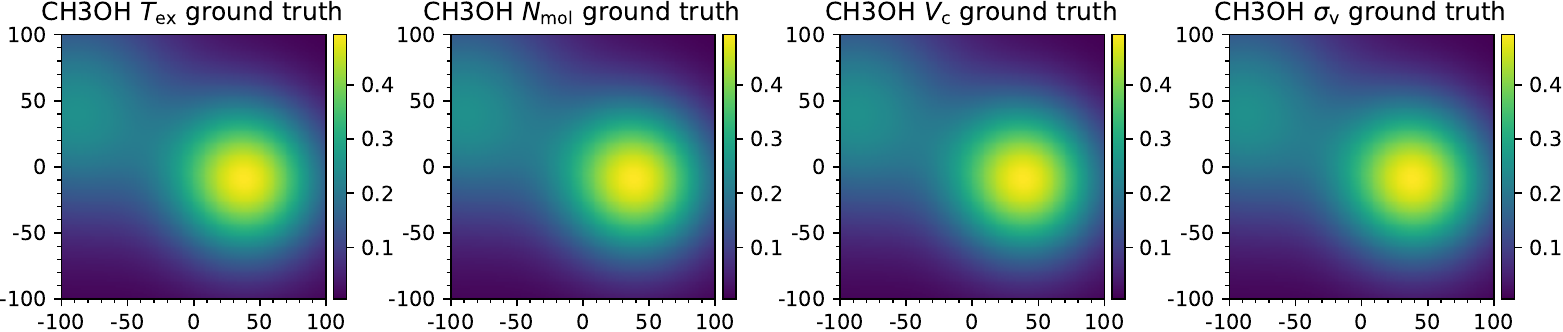}\\
  \includegraphics[scale=0.65]{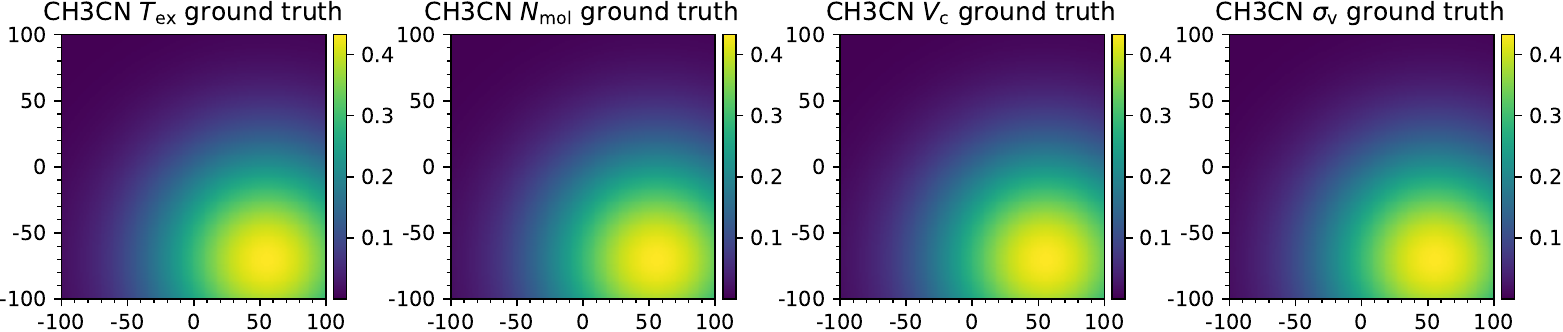}\\
   \includegraphics[scale=0.65]{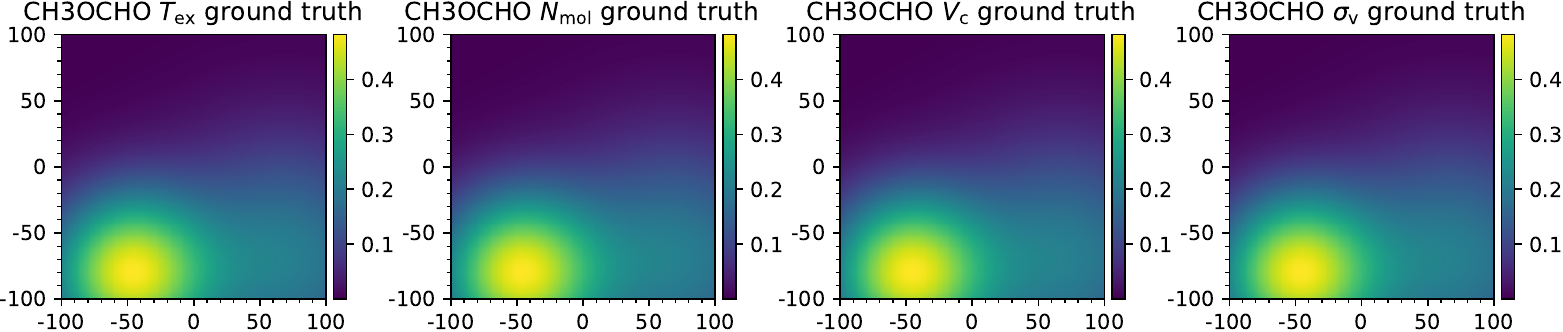}\\
\caption{Examples of 2D distributions of the ground-truth parameters, $T_{\mathrm{ex}}$, $N_{\mathrm{mol}}$, $V_{\mathrm{c}}$, and $\sigma_{\mathrm{v}}$ generated by a two-component normal distribution. The parameter ranges are all normalized to follow $\mathcal{N}$(0, 1).}
    \label{fig:syn_paras}
\end{figure*}

We created a synthetic datacube of 200$\times$200 grids in the position domain, having a spectral axis of 138 016 grids. To mock the structure of an observed source, we used a two-component normal distribution (multivariate normal distribution) for each set of parameters ($T_{\mathrm{ex}}$, $N_{\mathrm{mol}}$, $V_{\mathrm{c}}$, and $\sigma_{\mathrm{v}}$ for each molecule), for demonstration purposes. The peak or mean locations, and the variances of the two normal distributions are generated randomly for each molecule, with peak locations always within the position grid range. However, we emphasize that our proposed method, building GP models through BASQ, can handle more complex underlying distributions of the parameters due to the flexibility and expressive form of the kernel used. In particular, the radial basis function (RBF) kernel used in our experiment has the universal approximation property, ensuring that a GP with RBF kernel can asymptotically reconstruct any true function as the number of observations increases \citep{park1991universal}. However, this theory only guarantees the existence of such a reconstruction and does not imply that the RBF kernel always provides better sample efficiency for function approximation compared to other kernels in practice. Kernel selection significantly affects the convergence rate. For instance, the Matérn kernel is a suitable choice when the function is expected to be less smooth. As such, kernel design is as much an art as it is a science \citep{duvenaud2014automatic}, with the RBF kernel being a safe choice due to its universality. 
In reality, chemically related species, as well as molecules favoring similar excitation conditions, can share similar velocity parameters and excitation temperatures, rather than showing a random behavior of these parameter distributions. Yet, the randomly generated normal distribution of displaced peak locations and different variances should represent a broader range of cases, in general. Some examples of the ground-truth parameter maps used to generate the synthetic datacubes are shown in Fig. \ref{fig:syn_paras}. These parameter maps represent real parameter maps after standardization, where the original physical ranges for the four parameters $T_{\mathrm{ex}}$, $N_{\mathrm{mol}}$, $V_{\mathrm{c}}$, and $\sigma_{\mathrm{v}}$ are 50-500 K, 10$^{16}$-10$^{19}$ cm$^{-2}$, -3-3 km s$^{-1}$, and 0.5-10 km s$^{-1}$.

In the end, the intrinsic limit of inferring the parameter maps comes from the one-component LTE model, while the emission structure along the line-of-sight is a weighted average dependent on the density, temperature, and abundance profiles of a particular source. In any case, fast and robust inference of the parameter maps from one-component LTE models can serve as a foundation for further tailored analysis for individual sources and may represent the only feasible way to yield statistically significant results from a large-scale systematic survey.

\subsection{Cross-validation of the inferred parameters}
Cross-validation of the inferred parameters for individual locations against the ground-truth parameters is crucial for assessing the accuracy of the SVI step. The subsequent iterative construction of GP models heavily depends on both the accuracy and uncertainties of the SVI results from already-inferred locations. Figure \ref{fig:test_svi_92} illustrates the cross-validation results for 50 randomly selected initial locations. We find that $N_{\mathrm{mol}}$, $V_{\mathrm{c}}$, and $\sigma_{\mathrm{v}}$ show favorable consistency. However, a relatively larger scatter is observed in the low $T_{\mathrm{ex}}$ ground-truth regime. Primarily, these parameters are associated with highly degenerate line components of two molecules, HNCO and HN$^{13}$CO, and with another molecule CH$_{3}$COOH, which show mostly very weak emission lines, almost hidden in the strong line emission and close to the noise threshold. For the former two molecules, their similar frequency intervals between line components, combined with our randomly selected centroid velocities, result in severe merging of line components, rendering the problem unidentifiable. This degeneracy is partially reflected by the relatively large uncertainties associated with the fitted $T_{\mathrm{ex}}$. The severe line blending embedded in this synthetic spectrum setting is illustrated in Fig. \ref{fig:illu_line_bl}, showing a representative frequency range of 0.5 GHz. Figures \ref{fig:repre_sp1}-\ref{fig:repre_sp2} show an example of comparison between the synthetic spectrum and the SVI modeled spectrum. In reality, we do not expect the centroid velocities of most COMs to show large variations toward a single location. Ideally, one should have a measure of unidentifiability given the complexity and possible degeneracy of line-rich spectra for individual molecules. We plan to address this in future development, where such expert knowledge can be seamlessly incorporated into prior settings.

\subsection{BASQ inferred parameter maps through iterative refinement of GP}\label{sec:syn_test}
Following the steps elaborated in Sect. \ref{sec:gp}, through an iterative process of querying for the most informative locations (after the initial random query), refining the GP models to predict the output parameters and propose new query locations, we gradually reach a status where parameters at unknown locations do not improve the GP model predictions (the reduction of uncertainty) further. This can be defined by some stopping criterion, such as the RMSE reaching a plateau over the last batch of iterations.

\begin{figure*}
 \includegraphics[scale=0.65]{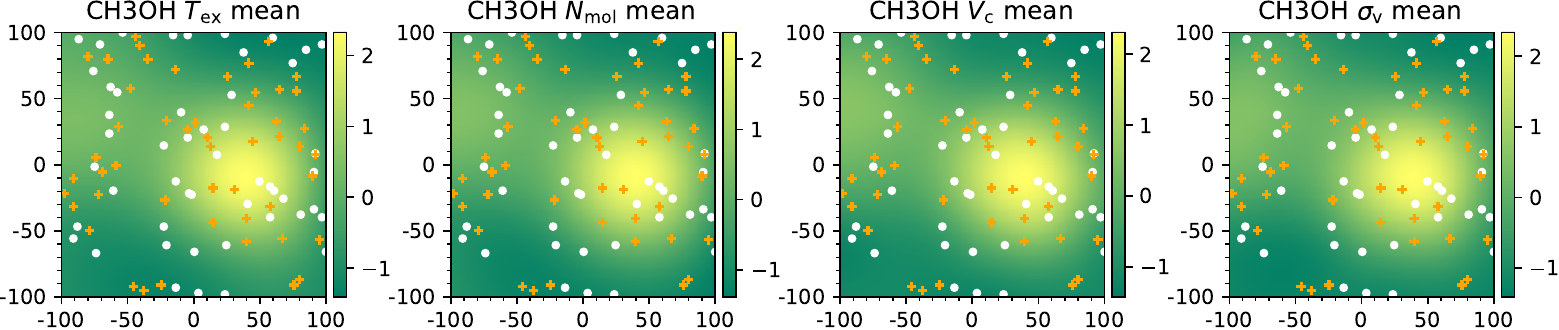}\\
 \hspace{-1.5cm}
  \includegraphics[scale=0.65]{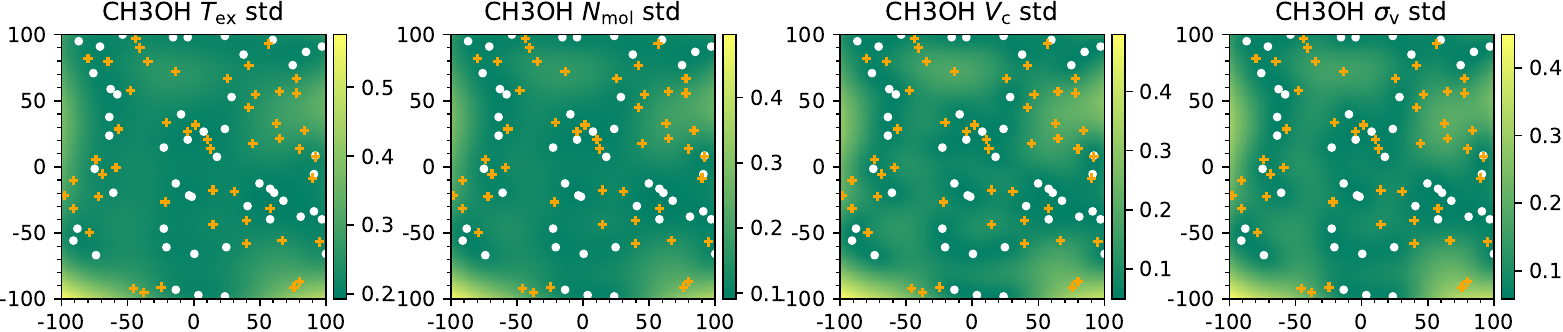}\\
   \includegraphics[scale=0.65]{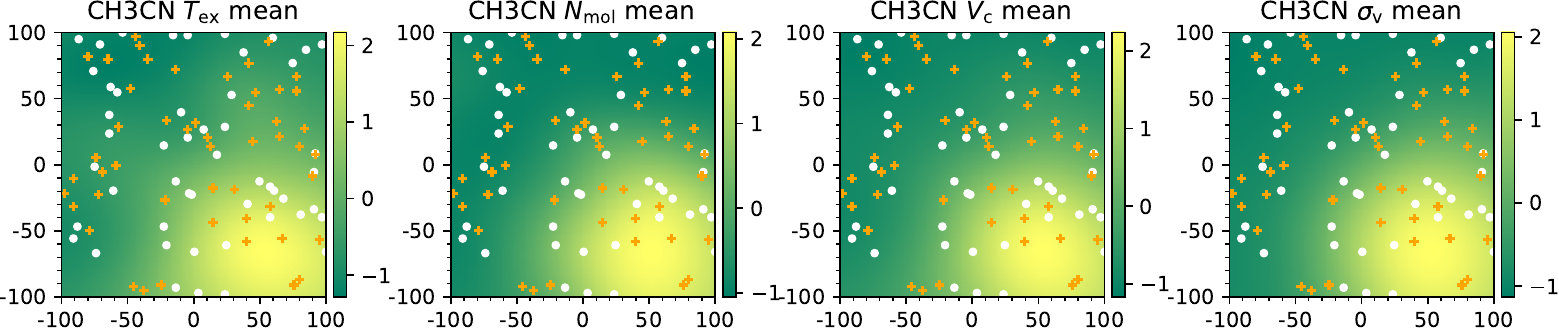}\\
   \includegraphics[scale=0.65]{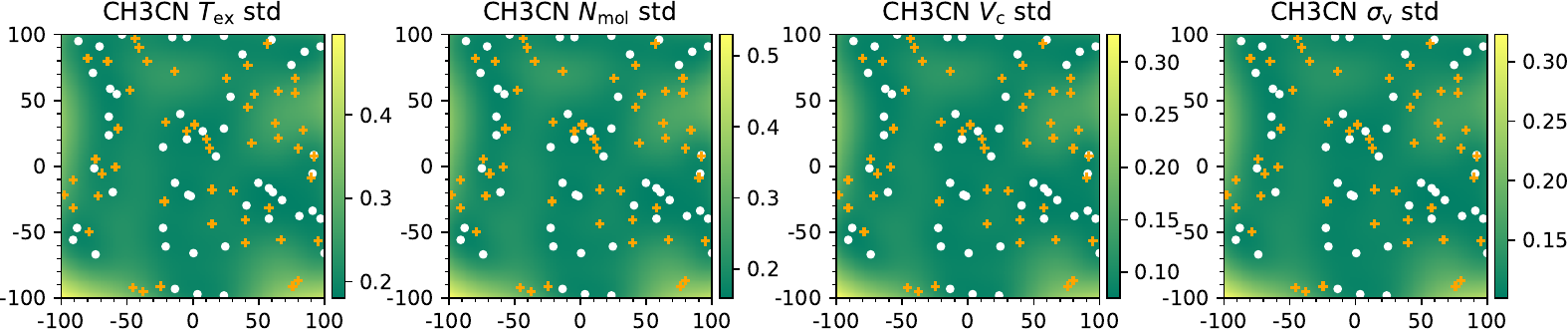}\\
   \includegraphics[scale=0.65]{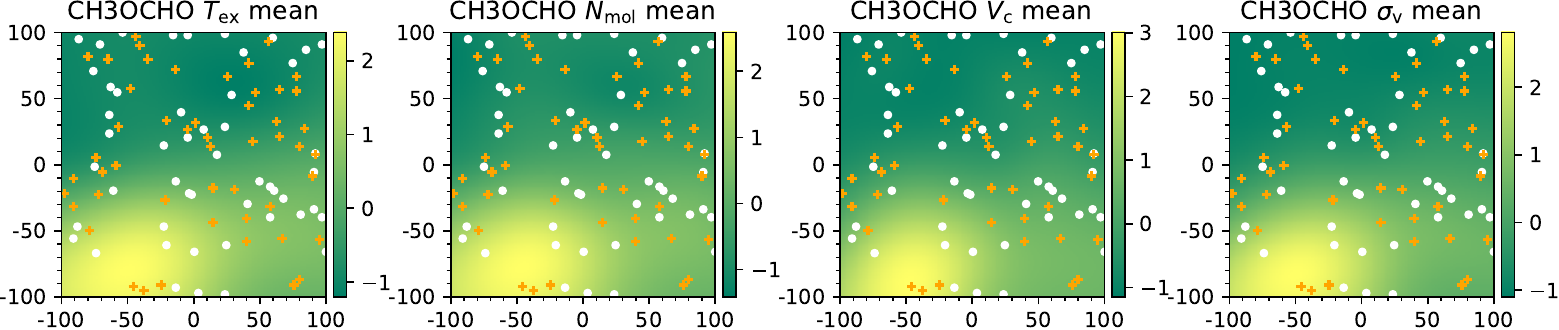}\\
   \includegraphics[scale=0.65]{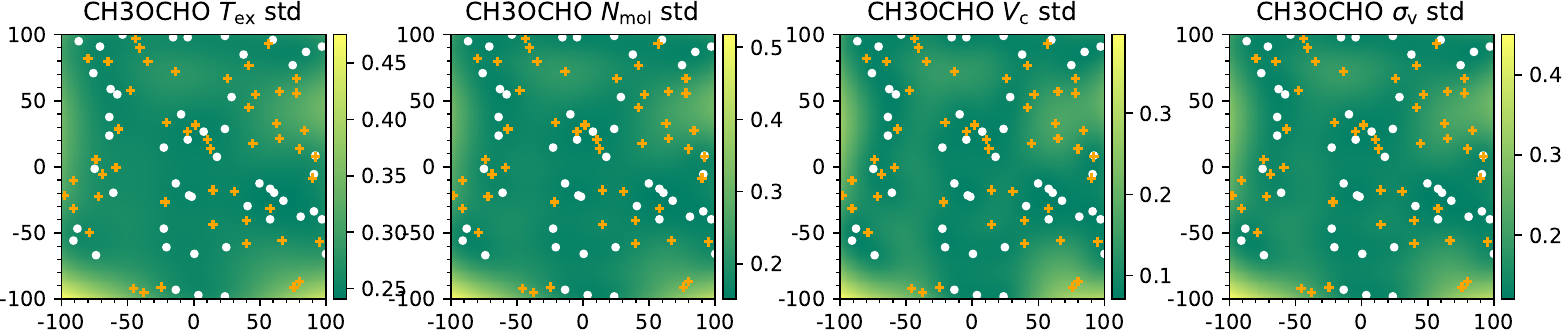}\\
      
\caption{First-iteration results of multi-output GP model predictions and predicted standard deviations for the four parameters, $T_{\mathrm{ex}}$, $N_{\mathrm{mol}}$, $V_{\mathrm{c}}$, and $\sigma_{\mathrm{v}}$, after SVI fitting to 50 initially randomly selected positions and model training. The white dots mark the locations where the SVI fitting is conducted, and the orange pluses indicate the proposed next-iteration locations to be queried, as suggested by the BASQ method.}
    \label{fig:basq_iter1}
\end{figure*}

\begin{figure*}
 \includegraphics[scale=0.75]{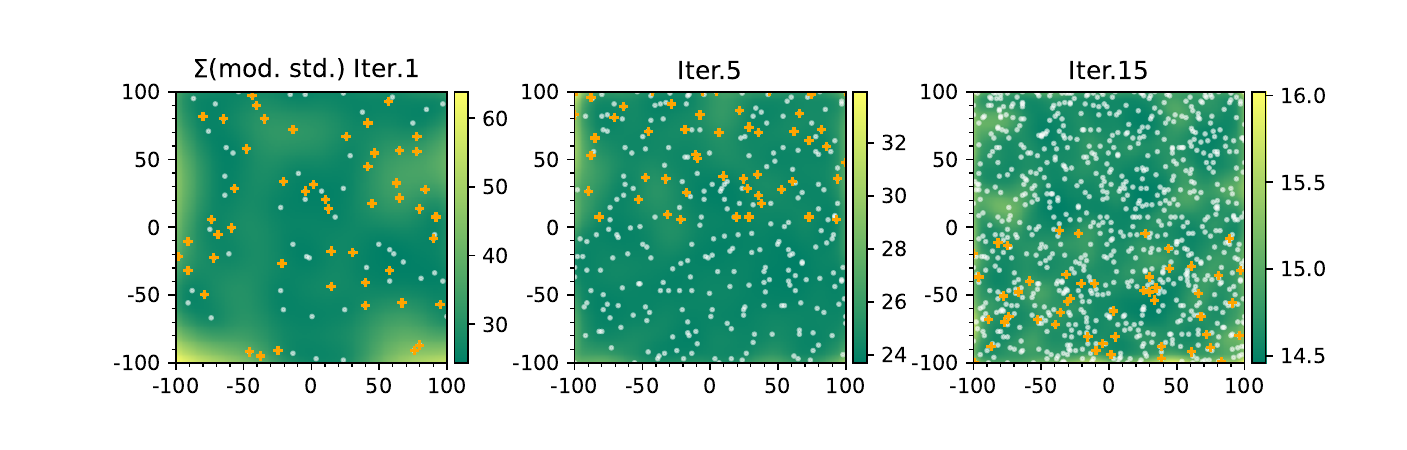}
 \caption{Sum of the GP model's predictive variances across all parameter maps, after the first, fifth, and fifteenth iteration of training. The white dots indicate the queried locations, while the orange pluses indicate the next-iteration locations proposed by the GP model.}
    \label{fig:basq_sum_var}
\end{figure*}

\begin{figure*}
 \includegraphics[scale=0.65]{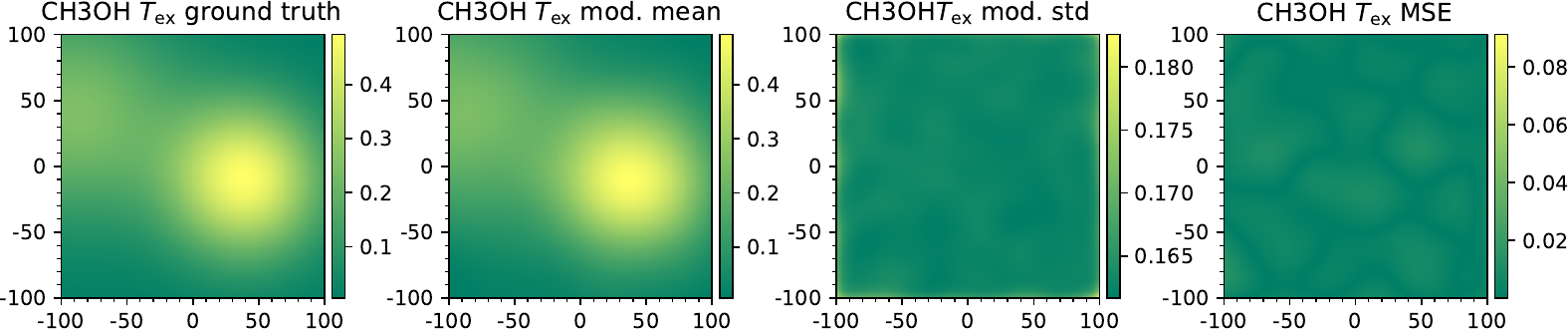}\\
 \hspace{-1.5cm}
  \includegraphics[scale=0.65]{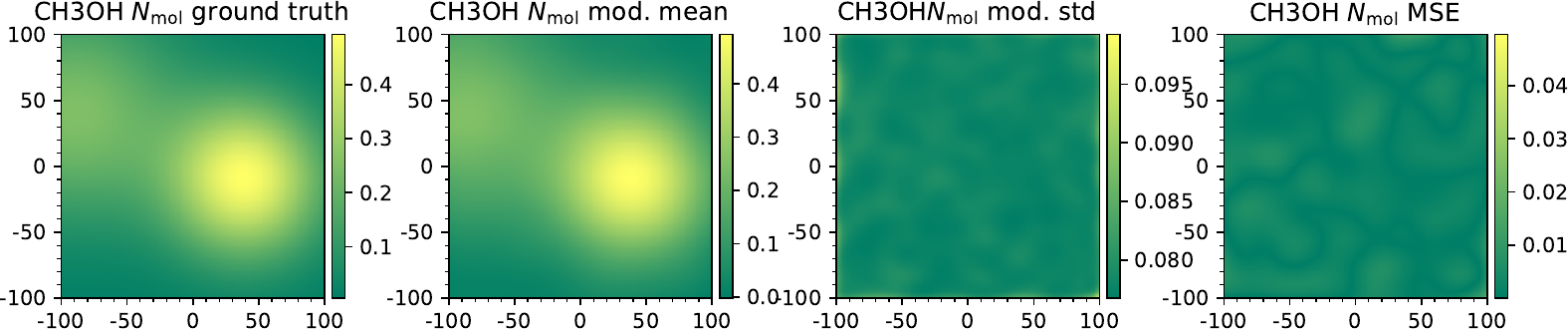}\\
   \includegraphics[scale=0.65]{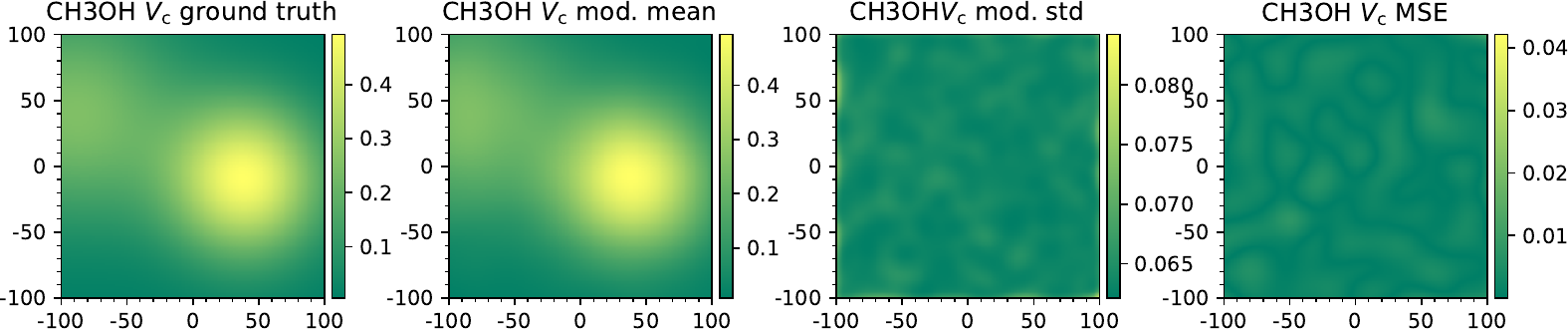}\\
   \includegraphics[scale=0.65]{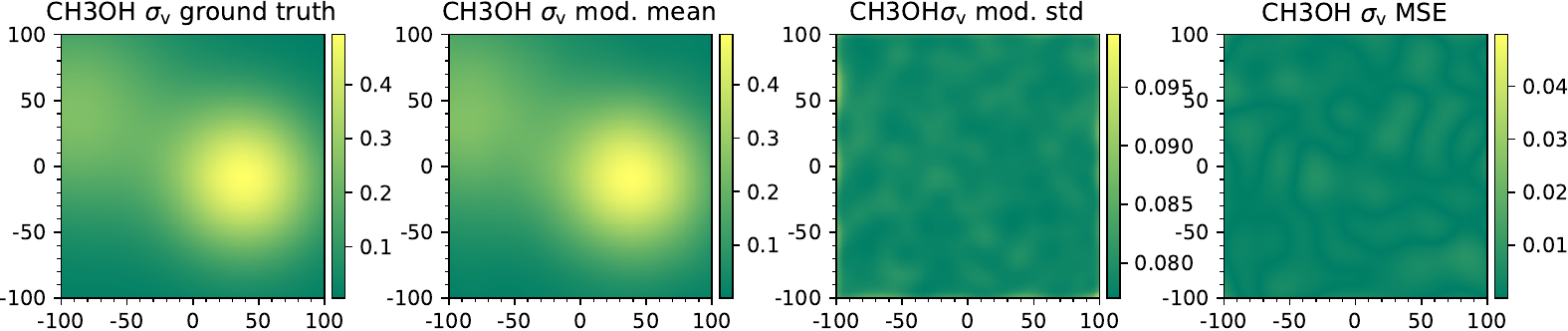}\\
  
\caption{Comparison between the ground truth and GP model-predicted parameter distributions after 18 iterations for CH$_{3}$OH. The two left panels show the ground truth and GP model-predicted mean distributions on the same color scale. The two right panels show the standard deviation of the model predictions and the MSE between the ground truth and model predictions. The maximum color scale for the MSE is set to 0.5 times the maximum value of the standard deviation of the model predictions.}
    \label{fig:compare_basq_ground_ch3oh}
\end{figure*}

Figures \ref{fig:basq_iter1}-\ref{fig:basq_iter15} illustrate the iterative process of GP model predictions based on queried locations and proposed new locations for subsequent rounds of queries. The figures depict the first, fifth, and fifteenth iterations, with a BASQ batch size of 50, proposing 50 locations per iteration for the next round of queries. The GP model predictions demonstrate progressive improvement, characterized by increasingly smooth predictions and lower standard deviations. Figures \ref{fig:compare_basq_ground_ch3oh}-\ref{fig:compare_basq_ground_ch3ocho} present comparisons between parametric ground truths and model predictions (using identical color scales) after the fifteenth iteration. These figures also display the standard deviation of the model alongside the difference between ground truths and model prediction, expressed as mean square error (MSE). The MSE values fall within the model prediction standard deviations and are approximately 10$\%$ of the ground-truth values, indicating robust model performance. Compared to pixel-by-pixel fitting, the BASQ framework saves computational time by $\sim$50 times in this case, by querying only the most informative locations (40,000 locations vs. 1000 locations).

\section{Benchmarking with PILS results: Molecular column densities in IRAS16293B at 0.5$''$ offset position}
We benchmarked the SVI step with the published result from the PILS survey (\citealt{Jorgensen16}). The column densities for more than 130 molecules have been reported toward IRAS16293B, some values of which are fit by fixed $T_{\mathrm{ex}}$ temperatures of 125 K and 300 K. Also, column densities for molecules that have mostly optically thick lines are derived using fixed isotope ratios based on the column densities of the rarer isotopologues. We excluded these molecules with indirectly fitted column densities in the comparison. We also ignored molecules that only have less than three transitions in the frequency coverage due to the larger uncertainties in constraining the LTE model.

\begin{figure}[htb]
\centering
\includegraphics[scale=0.5]{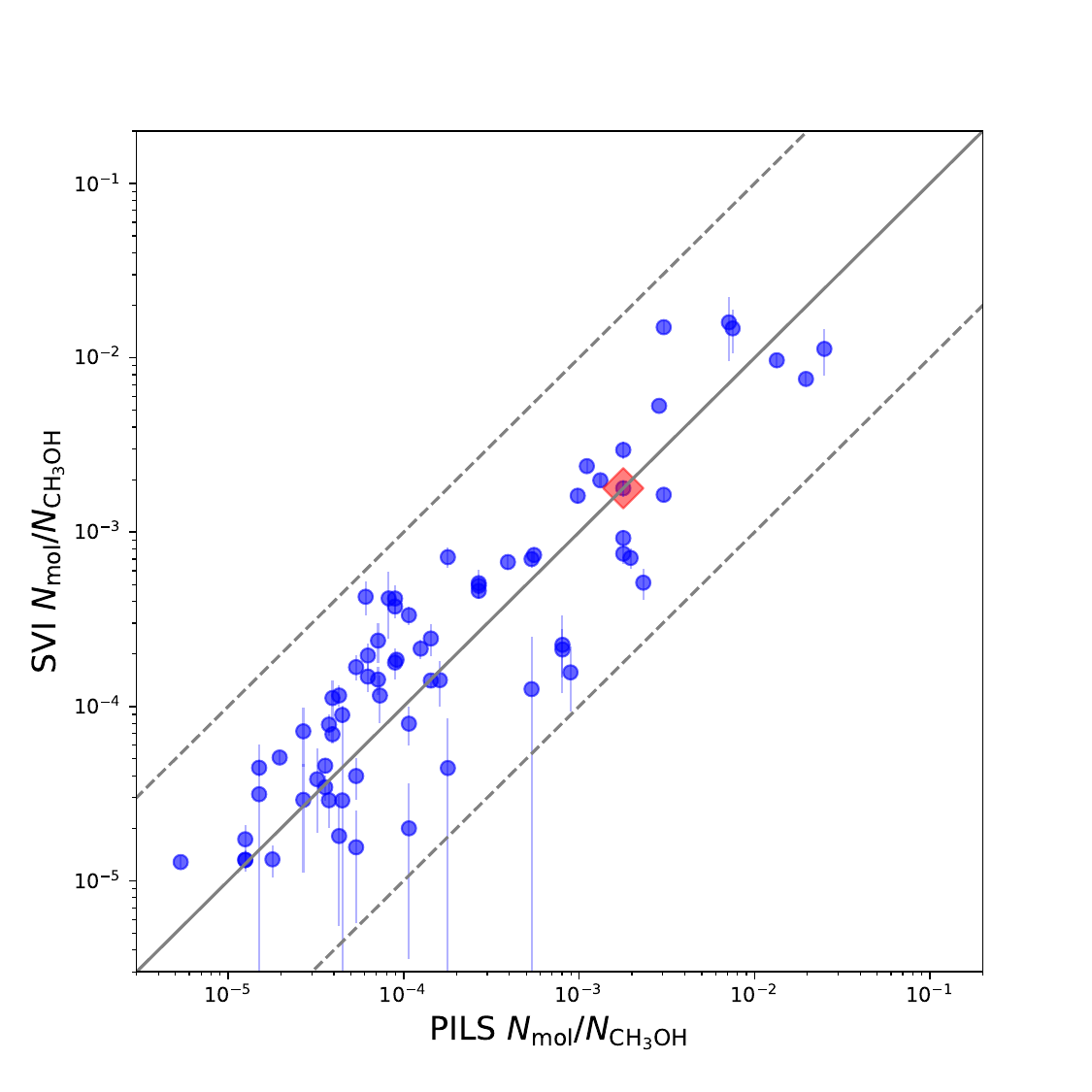}
\caption{Cross-validation of the SVI-fitted column densities with the PILS results from the literature. The red square indicates the location of CH$_{3}$$^{18}$OH, from which the column density of CH$_{3}$OH is derived using a fixed $^{18}$O/$^{16}$O ratio. The solid line indicates the equal line, and the dashed lines show factors of 0.2 and 5 difference.}
\label{fig:bm_pils_noTexfix}
\end{figure}

The comparison between SVI-derived column densities and results published in the PILS survey is presented in Fig. \ref{fig:bm_pils_noTexfix}. It is important to note that the PILS survey column densities often focus on specific molecules and sometimes rely on relatively isolated lines for a particular molecule (and its isotopologues), rather than employing a global optimized fitting approach using a comprehensive list of molecules (Table \ref{tab:mollist}). Consequently, differences in derived column densities are expected. While it is challenging to directly compare the individual fitting errors from the PILS survey with our SVI global fitting results, our comparison reveals that data points showing large differences are correctly characterized by larger standard deviations in the SVI fitting (as shown by error bars in Fig. \ref{fig:bm_pils_noTexfix}). Although expert-knowledge-based fine-tuning for individual molecules may sometimes yield more robust results than global fitting, the latter approach requires careful consideration of potential line blending effects. Nonetheless, SVI global fitting results can provide a valuable foundation for further manual tuning and enable more targeted MCMC fitting for specific parameter spaces, depending on available computing resources.

\section{Caveats and outlook}
There are basically three blocks in our developed framework, which are separable for alternative methods or algorithms serving similar purposes at each level. In particular, the third step, building GP models with the BASQ technique iteratively, highly benefits from the robustness of the inferred (fitted) results and the associated uncertainties in the second step. The drawbacks of VI, compared to an MCMC sampler such as NUTS and other state-of-the-art algorithms, are broadly discussed in the literature (\citealt{Blei16}; \citealt{Hoffman20}). Nevertheless, VI itself is a rapidly developing field (e.g., \citealt{geovi}), and hybrid methods that combine VI and MCMC (\citealt{Salimans14}; \citealt{Geffner21}) have been proposed to leverage the strength of both approaches. On the other hand, the third step itself, by nature, has a wider application realm in speeding up reconstructing parameter maps from observed datacubes based on arbitrary models, which is particularly relevant for large-scale, high-resolution (both in frequency and spatial resolution) surveys. However, BASQ does not support asynchronous parallelization, where SVI fitting results arrive at different times, requiring dynamic reallocation of the batching budget. In such cases, alternative approaches as Thompson sampling \citep{kandasamy2018parallelised} could be considered.

The machinery developed for our spectral fitting tool will be essential for the ALMA Wideband Sensitivity Upgrade (WSU), which enables significantly more sensitive and broader-bandwidth spectroscopic observations. As the WSU introduces next-generation receivers, digitizers, and a powerful new correlator system, the resulting datasets will be larger and more complex than ever before. Our tool is designed to handle the challenges of wideband, high-sensitivity spectral surveys, making it a critical component for extracting accurate physical and chemical information from the upgraded facility’s observations.

\section{Conclusions}
In conclusion, this study presents a novel and efficient workflow for estimating molecular parameter maps in hot cores and hot corinos, addressing the challenges posed by line-rich spectra and severe line blending. Our approach combines SVI with Bayesian active learning and parallelization strategies to achieve accurate and computationally efficient results. The key findings and implications of our work are as follows:
   \begin{enumerate}
      \item The SVI provides reliable estimates of molecular parameters at individual positions, effectively accounting for line blending and unknown contaminant features.
      \item The combination of Bayesian active learning and parallelization significantly reduces the computational cost of generating high-fidelity molecular parameter maps compared to exhaustive pixel-by-pixel fitting.
      \item Our workflow is capable of handling hundreds of molecules simultaneously, making it suitable for wideband spectral datacubes of complex astrochemical environments.
The continuous, high-angular-resolution maps generated by our method offer detailed insights into the spatial distributions of COMs in hot cores and hot corinos.
\item Benchmarking against published column densities from the PILS survey demonstrates the accuracy and reliability of our approach.
   \end{enumerate}
This work represents a significant step forward in the analysis of wideband line-rich spectral data from hot cores and hot corinos. By enabling a more efficient and accurate estimation of molecular parameter distribution through a state-of-the-art active learning approach, our method paves the way for a deeper understanding of astrochemical pathways, especially when probing the spatial distribution and variations of molecules is important. Future applications of this workflow to large observational datasets promise to yield valuable insights into the chemical complexity and evolution of star-forming regions.

%\bibliography{ref}

\section{Data availability}
A minimal working example of the complete workflow, along with a synthetic dataset, is available at \url{https://github.com/yxlinaqua/BaSIL/}. Ongoing development, including applications to observational data, will be presented in future work.

The workflow makes use of the following Python packages: \texttt{astropy}, \texttt{botorch}, \texttt{gpytorch}, \texttt{jax}, \texttt{joblib}, \texttt{linear\_operator}, \texttt{matplotlib}, \texttt{numpy}, \texttt{numpyro}, \texttt{optax}, \texttt{scipy}, \texttt{torch}.

\begin{acknowledgements}
    V. Eberle acknowledges support for this research through the project Universal Bayesian
Imaging Kit (UBIK, F\"orderkennzeichen 50OO2103) funded by the Deutsches Zentrum für Luft- und Raumfahrt e.V. (DLR). Masaki Adachi was supported by the Clarendon Fund, the Oxford Kobe Scholarship, the Watanabe Foundation, and Toyota Motor Corporation. Y. Lin gratefully acknowledges the 2023 Carl-Zeiss-Stiftung Summer School on Scientific Machine Learning for Astrophysics for the stimulating experience, and also thanks Antoine Alaguero for kindly providing their simulation data run for V892 Tau.
\end{acknowledgements}

\begin{appendix}
\section{Supplementary explanations of key concepts and methods}

\subsection{Gaussian process}
\paragraph{Definitions}
Let $(\Omega, \mathcal{F}, \mathbb{P})$ be a probability space and $\mathcal{X} \subseteq \mathbb{R}^{d}$ represent the input domain. A GP \citep{stein1999interpolation, Rasmussen} is a stochastic process denoted as $g: \mathcal{X} \times \Omega \rightarrow \mathbb{R}$. The characteristics of this process are defined by its mean function $m: \mathcal{X} \rightarrow \mathbb{R}$, where $m(x) = \mathbb{E}[g(x, \cdot)]$, and its covariance function $K: \mathcal{X} \times \mathcal{X} \rightarrow \mathbb{R}$, given by $K(x, x^\prime) = \mathbb{E}[(g(x, \cdot) - m(x))(g(x^\prime, \cdot) - m(x^\prime))]$. The covariance function is both symmetric ($K(x,x^\prime)=K(x^\prime, x)$ for all $x, x^\prime \in \mathcal{X}$) and positive definite (for any $t \in \mathbb{N}$, $\{ a_i \}_{i=1}^t \subset \mathbb{R}$, $\{ x_i \}_{i=1}^t \subset \mathcal{X}$, it holds that $\sum^t_{i,j=1} a_i a_j K(x_i, x_j) \geq 0$). Functions meeting these criteria are referred to as kernels. A GP defines a probability measure over functions and can be conditioned on data in a closed form for conjugate likelihood scenarios.

In a regression context, we assume the labels follow the model $y = f(x) + \epsilon$, where $f$ is the underlying function to be estimated, and $\epsilon \sim \mathcal{N}(0, \sigma^2)$ represents independent and identically distributed (i.i.d.) zero-mean Gaussian noise with variance $\sigma^2$. Given a labeled dataset $\mathcal{D}t = \{ x_i, y_i \}_{i=1}^t := (\textbf{X}_t, \textbf{Y}_t)$ and the corresponding covariance matrix $\textbf{K}_{XX} = (K(x_i, x^\prime_j))_{1\leq i,j\leq t} \in \mathbb{R}^{t \times t}$, also known as \emph{Gram matrix}, the GP regression model conditioned on this data is expressed as $f \mid D_t \sim \mathcal{GP}(m_t, C_t)$, where
\begin{align*}
    \begin{split}
    m_t(x) &= m(x) + K(x, \textbf{X}_t) (\textbf{K}_{XX} + \sigma^2 \textbf{I}_{t \times t})^{-1}(\textbf{Y}_t -m(\textbf{X}_t)),\\
    C_t(x, x^\prime) &= K(x, x^\prime) - K(x, \textbf{X}_t) (\textbf{K}_{XX} + \sigma^2 \textbf{I}_{t \times t})^{-1} K(\textbf{X}_t, x^\prime),
    \end{split}
\end{align*}
$m_t(\cdot)$ and $C_t(\cdot, \cdot)$ are the mean and covariance functions of the GP posterior predictive distribution conditioned on $t$-th dataset $D_t$, and $\textbf{I}_{t \times t}$ is an identity matrix of size $t$.

\paragraph{Hyperparameter optimization}
A GP is characterized by its kernel $K$, and the kernel $K$ has hyperparameters $\theta$. Hyperparameters are optimized by maximizing the marginal likelihood with
\begin{align*}
    \begin{split}
    \mathcal{L}(\theta) := \mathcal{N}(\textbf{Y}_t; m_t(\textbf{X}_t), C_t(\textbf{X}_t, \textbf{X}_t)).
    \end{split}
\end{align*}
In GPyTorch, it adopts the efficient computation via Cholesky decomposition (see details in \cite{Rasmussen}).

\paragraph{Kronecker structure in multi-output GP models}
We employed a multi-output GP with an intrinsic coregionalization model (ICM) kernel to model all outputs for this problem \cite{maddox2021bayesian}. For $M$ tasks (outputs, or SVI fitting parameters) and $t$ data points, it is assumed that the responses, $Y \in \mathbb{R}^{t \times M}$, follow the distribution $\text{vec}(Y) \sim \mathcal{N}(f,D)$, with $f \sim \mathcal{GP}(\mu_{\theta}, K_{XX} \otimes K_T)$, where $D$ represents a (diagonal) noise term.

\subsection{Bayesian quadrature}
Assume a function $f_t$ is modeled by a GP, $f_t \sim \mathcal{GP}(m_t, C_t)$. Our goal is to estimate the expectation of the function $\hat{Z} := \int f(x) \, \text{d} \pi(x)$. This approach is known as Bayesian Quadrature (BQ) \citep{o1991bayes, hennig2022probabilistic}, and the integral estimates are given by
\begin{subequations}
\begin{align}
    \mathbb{E}_{f_t \sim \mathcal{GP}(m_t, C_t)} [ \hat{Z}] 
    &= \int m_t(x)\, \text{d} \pi(x)
    = \boldsymbol{z}^\top_t (\textbf{K}_{XX} + \sigma^2 \textbf{I}_{t \times t})^{-1} \textbf{Y}_t,
\label{eq:bq_mean}\\
    \mathbb{V}_{f_t \sim \mathcal{GP}(m_t, C_t)} [ \hat{Z} ]
    &= \int C_t(x, x^\prime) \, \text{d} \pi(x) \, \text{d} \pi(x^\prime)
    = z^\prime_t - \boldsymbol{z}^\top_t (\textbf{K}_{XX} + \sigma^2 \textbf{I}_{t \times t})^{-1} \boldsymbol{z}_t,
\label{eq:bq_var}
\end{align}
\end{subequations}
where \(\boldsymbol{z}_t := \int K(x, \textbf{X}_t) \, \text{d} \pi(x)\) and \(z^\prime_t := \int K(x, x^\prime) \, \text{d} \pi(x) \, \text{d} \pi(x^\prime)\) represent the kernel mean and variance, respectively. To improve integration accuracy, it is essential to minimize the uncertainty in the integral estimation as described in Eq.(\ref{eq:bq_var}). Consequently, Eq.(\ref{eq:bq_var}) serves as the metric for assessing the reduction in integral variance, which has been utilized as the Acquisition Function (AF) for BQ \citep{rasmussen2003bayesian, osborne2012active}.

\section{Benchmark BASQ with parameter maps from numerical simulations of a binary system}\label{app:benchmark_rmse}
We use the physical parameters of hydrogen column density $N_{\mathrm{H_{2}}}$, dust temperature $T_{\mathrm{dust}}$, (line-of-sight) velocity line-width $\sigma_{V}$ and centroid velocity $V_{\mathrm{c}}$ from a state-of-the-art hydrodynamical simulation of a binary system (V892 Tau, \citealt{Alaguero24}) as the mock parameter maps we aim to reconstruct from the BASQ step. These maps contain characteristic localized structures (spiral arms) as well as large-scale structures, and therefore represent a more realistic picture of a complex system, showing more resemblance to the real observations than a simple multivariate distribution. In the purpose of fitting a parameter map of such properties, for this GP model, we use a combined RBF kernel plus a Matérn kernel (see, e.g., \citealt{Czekala15}, \citealt{Meech22}), in order to capture both the smoothness and the roughness of the underlying parameter maps.  The fine tuning of GP model parameters is beyond the scope of this manuscript, yet we emphasize the importance of experimenting the kernel definition when employing our workflow. 

To gauge the performance of the map reconstruction step (the third step in our workflow), we isolate it to conduct an independent experiment and benchmark it against a nearest neighbor (NN) interpolation method to assess the efficiency gained by BASQ. In order to simulate the SVI query results (second step in our workflow), we add random noise to the ground-truth maps, and use these "noisy" fitting results to train and iterate the GP model refinement with the aid of BASQ. We use variations of RMSE as a function of iteration time for the benchmark purpose. 

\begin{figure*}
    \includegraphics[width=0.95\linewidth]{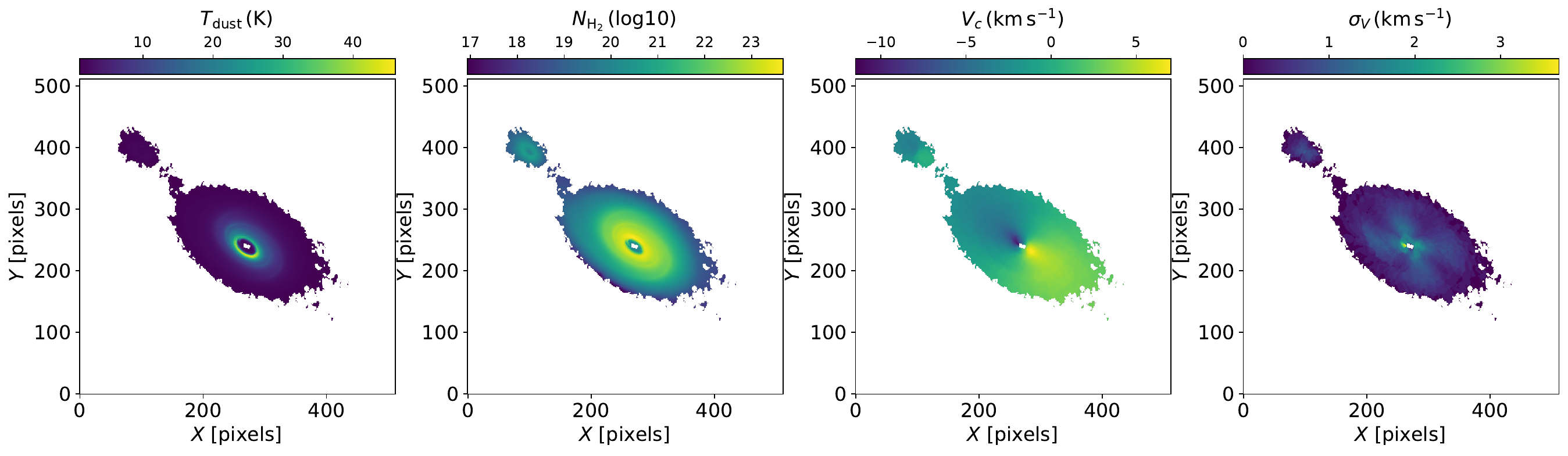}
    \label{fig:simu_4}
    \caption{Physical parameters of $T_{\mathrm{dust}}$, $N_{\mathrm{H_{2}}}$, $V_{\mathrm{c}}$ and $\sigma_{V}$ from simulations in \citealt{Alaguero24} of a binary system. Shown are the ground-truth parameter maps used in Appendix \ref{app:benchmark_rmse}. }
\end{figure*}

\begin{figure}
\includegraphics[width=0.95\linewidth]{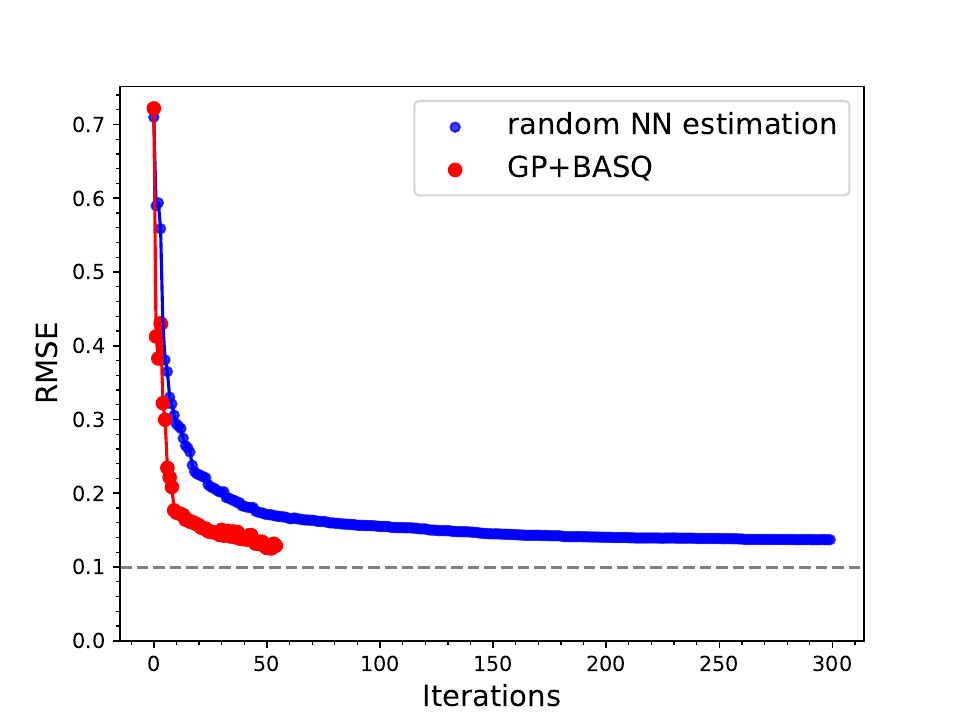}
\label{fig:rmse_benchmark_basq}
\caption{RMSE as a function of iteration number for NN estimation and the GP model plus BASQ algorithm in reconstructing the parameter maps in Fig. \ref{fig:simu_4}. The horizontal dashed line indicates the minimum achievable RMSE, resulting from noisy fitting at individual locations.}
\end{figure}

The four parameter maps from simulations are shown in Fig. \ref{fig:simu_4}, of dimension 512$\times$512 each. For the NN interpolation, we use the same batch size of 50 to update the training data from the mock SVI step, but with randomly selected query locations, in contrast to the BASQ step, where locations that are predicted to be more informative in reconstructing the parameter maps from the GP model is used for query. The RMSE plot is shown in Fig. \ref{fig:rmse_benchmark_basq}. Since we use the noisy SVI result for constructing the GP models, there is a best possible minimum defined by the added noise, which is included as a horizontal line in Fig. \ref{fig:rmse_benchmark_basq}. It can be clearly seen that the RMSE of Gaussian model training aided by BASQ arrives at a stable value with much smaller iteration number, than a random NN interpolation method. Since the bottleneck in the whole workflow is the second step--fitting hundreds of parameters at a single location, our raised algorithm maximally reduce the computational time by reducing the number of locations to fit with, while ensuring a smooth parameter map reconstruction, taking care of both the parameter correlations as well as the possible complex emission structure of a system. This reflects both the efficiency of the active learning framework of BASQ and the versatility of the GP models, which are particularly relevant for a quick yet robust parameter visualization for interpreting the line-rich datacubes.

\section{Additional figures and tables}

\begin{figure*}
    \includegraphics[width=\linewidth]{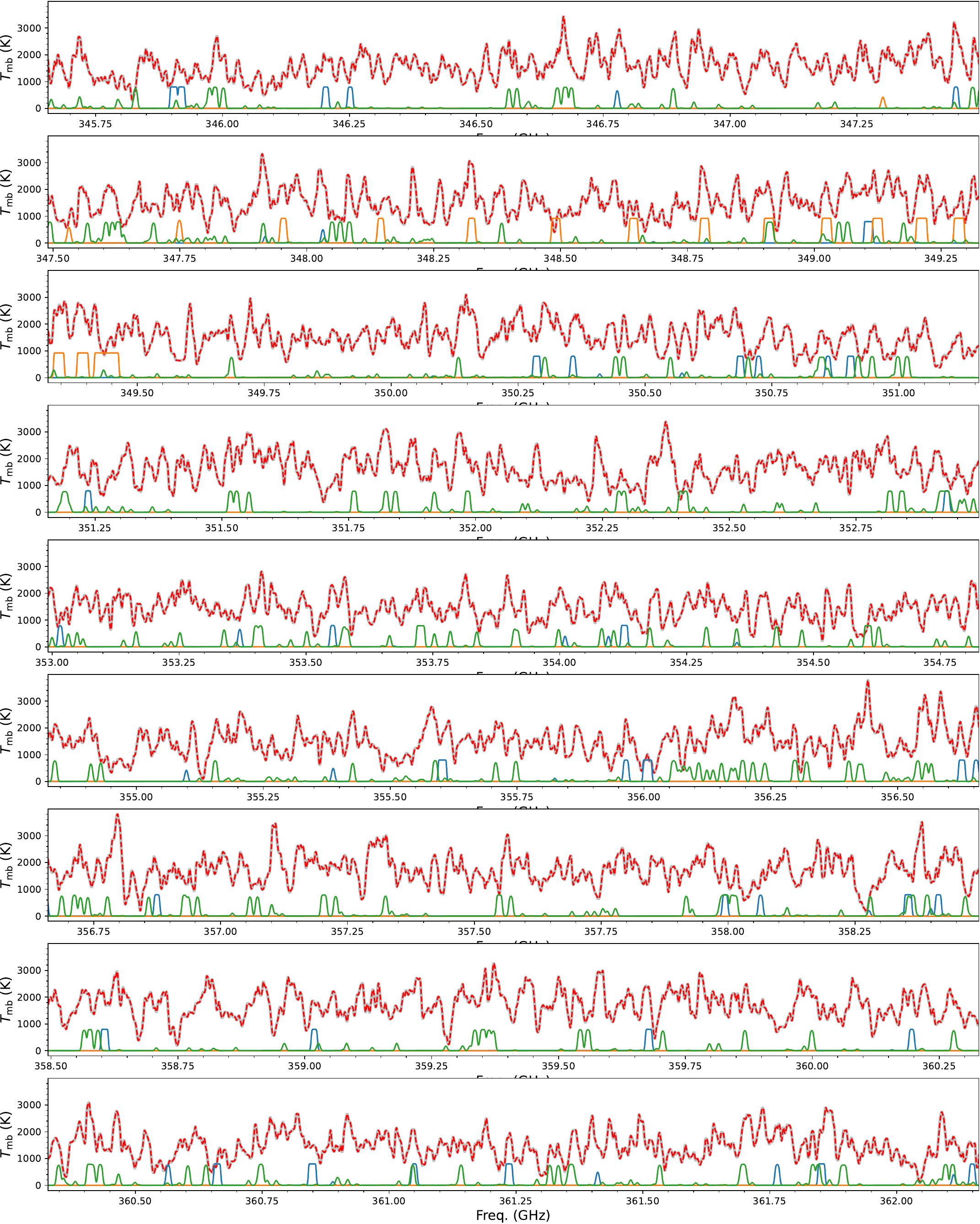}
    \caption{Same as Fig. \ref{fig:repre_sp1}, extended frequency range.}
    \label{fig:repre_sp2}

\end{figure*}

\begin{figure*}
 \includegraphics[scale=0.65]{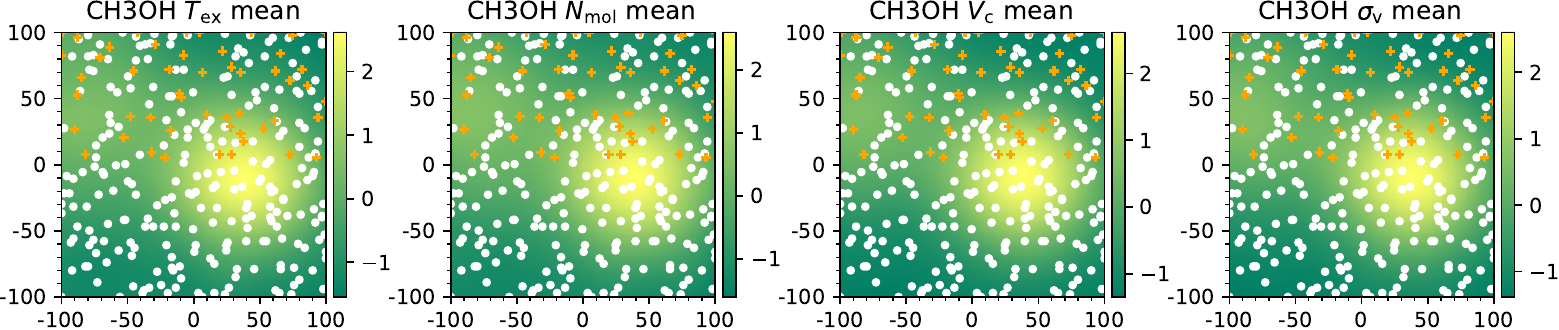}\\
 \hspace{-1.5cm}
  \includegraphics[scale=0.65]{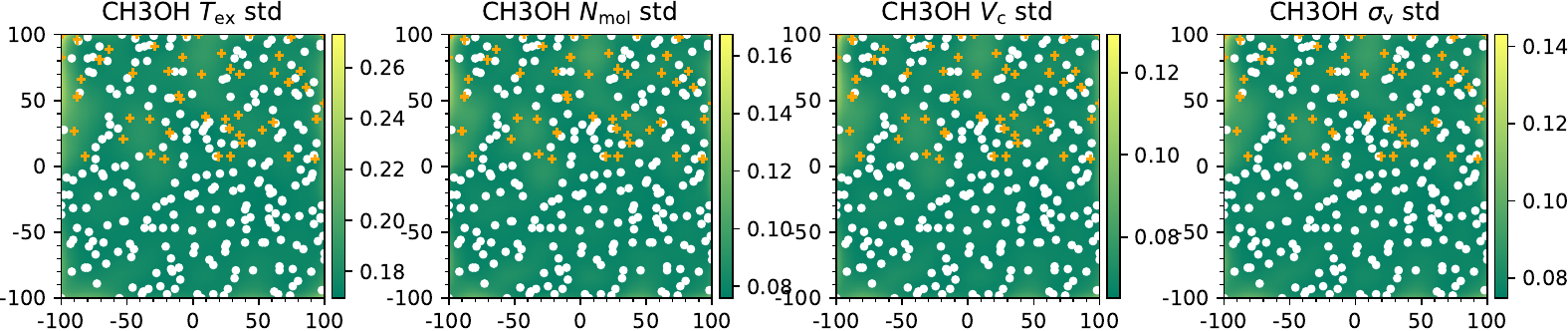}\\
   \includegraphics[scale=0.65]{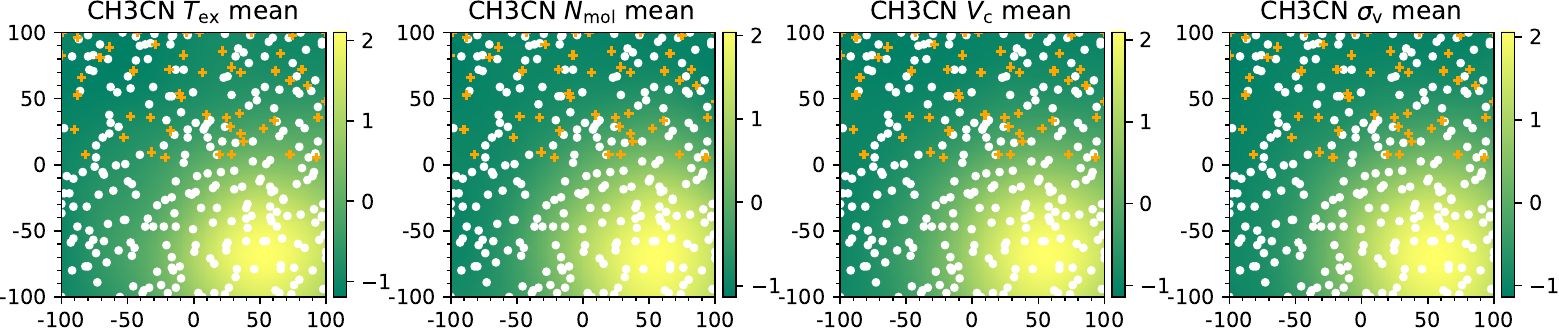}\\
   \includegraphics[scale=0.65]{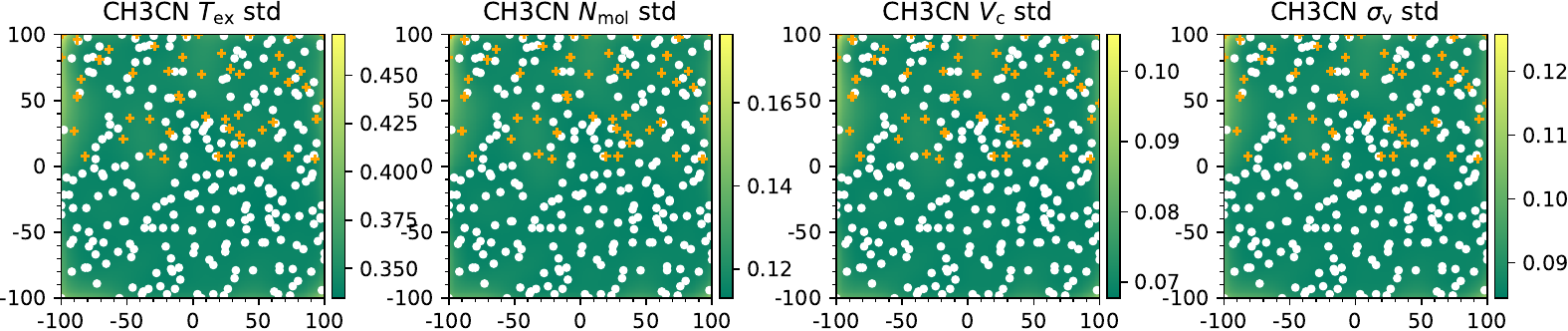}\\
   \includegraphics[scale=0.65]{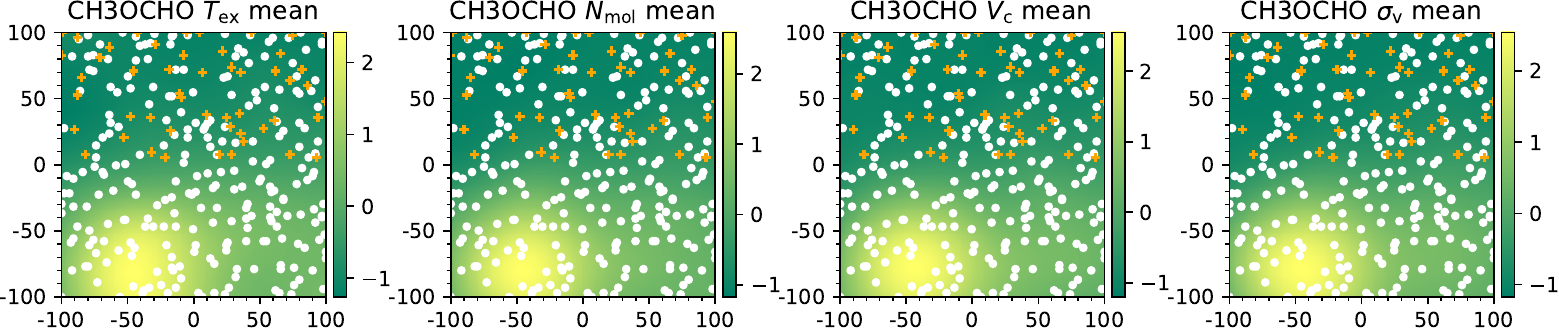}\\
   \includegraphics[scale=0.65]{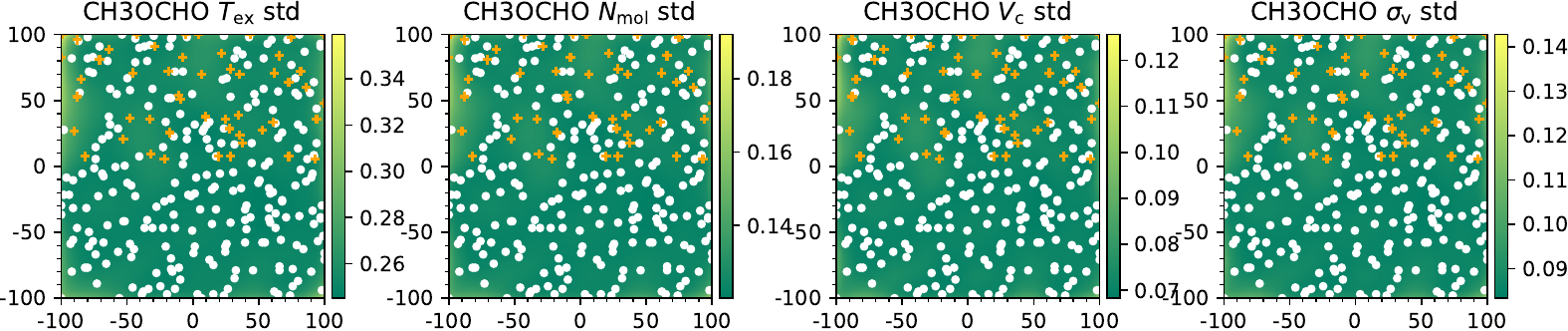}\\
      
\caption{Same as Fig. \ref{fig:basq_iter1}, but for model results after five iterations. Compared to the first model-predicted standard deviations of each parameter in Fig. \ref{fig:basq_iter1}, the values of the updated standard deviation are significantly suppressed.}
    \label{fig:basq_iter5}
\end{figure*}

\begin{figure*}
 \includegraphics[scale=0.65]{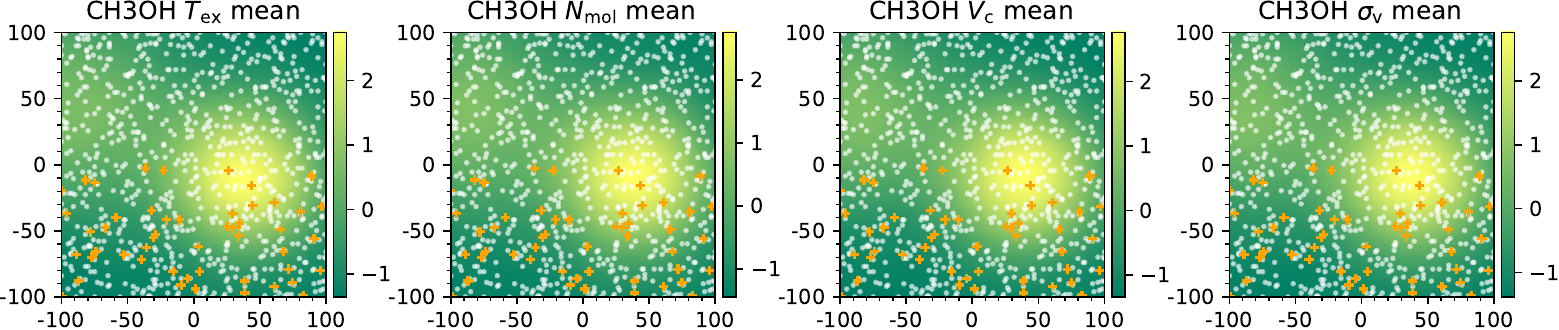}\\
 \hspace{-1.5cm}
  \includegraphics[scale=0.65]{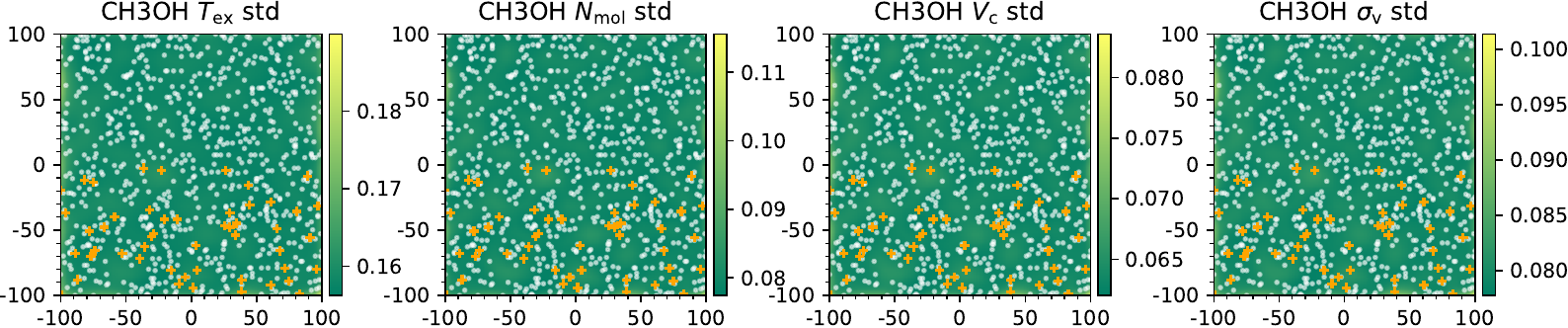}\\
   \includegraphics[scale=0.65]{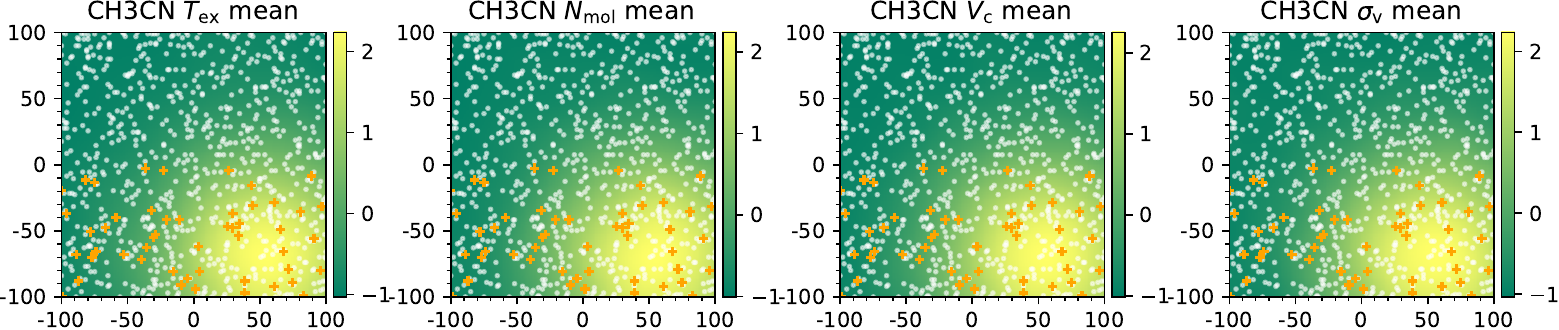}\\
   \includegraphics[scale=0.65]{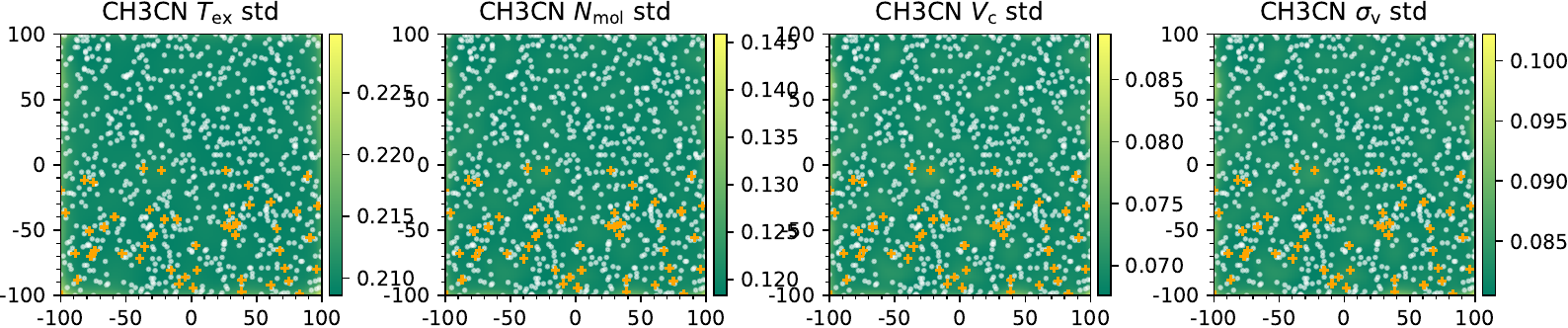}\\
   \includegraphics[scale=0.65]{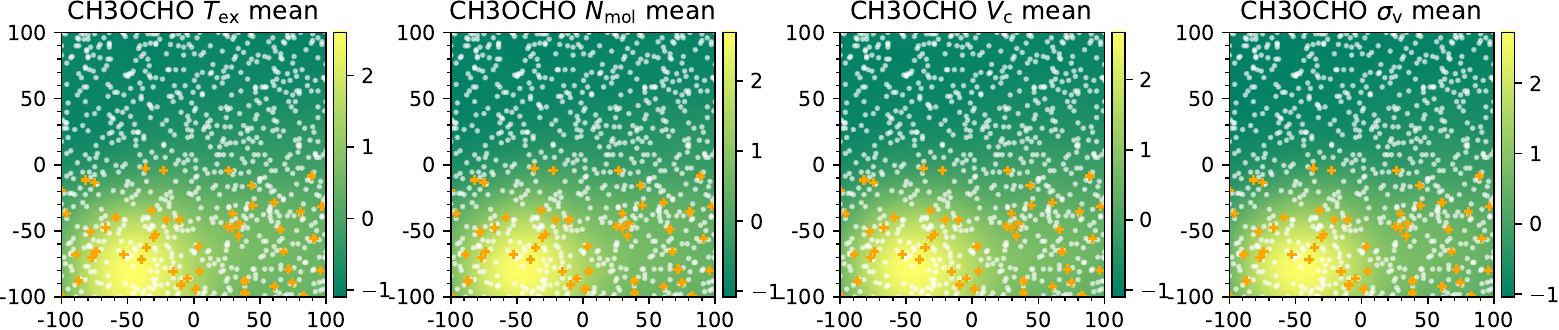}\\
   \includegraphics[scale=0.65]{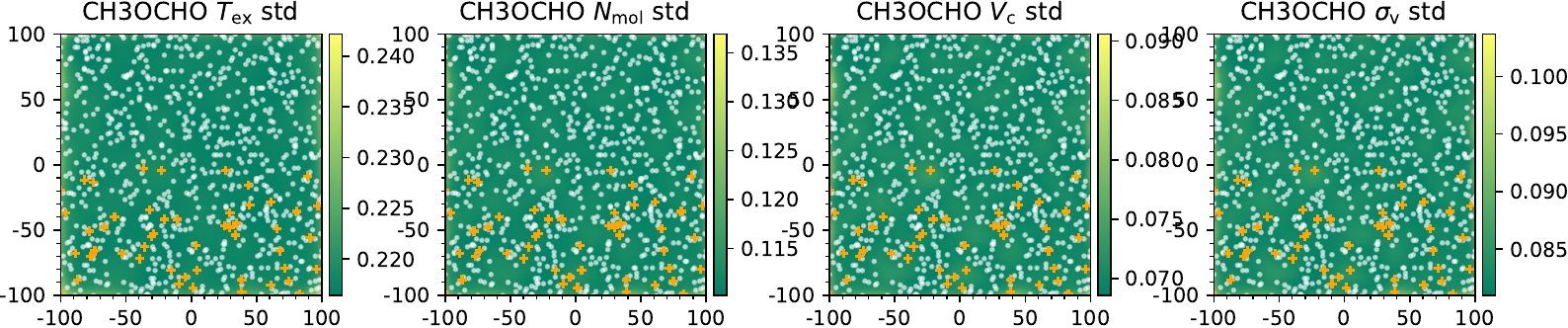}\\
      
\caption{Same as Fig. \ref{fig:basq_iter1}, but for model results after 15 iterations. The markers of queried data points are reduced in size to facilitate reading the color scale. The model predictions are smoother with standard deviation further suppressed, compared to results after five iterations, as shown in Fig. \ref{fig:basq_iter5}. }
    \label{fig:basq_iter15}
\end{figure*}

\begin{figure*}
 \includegraphics[scale=0.65]{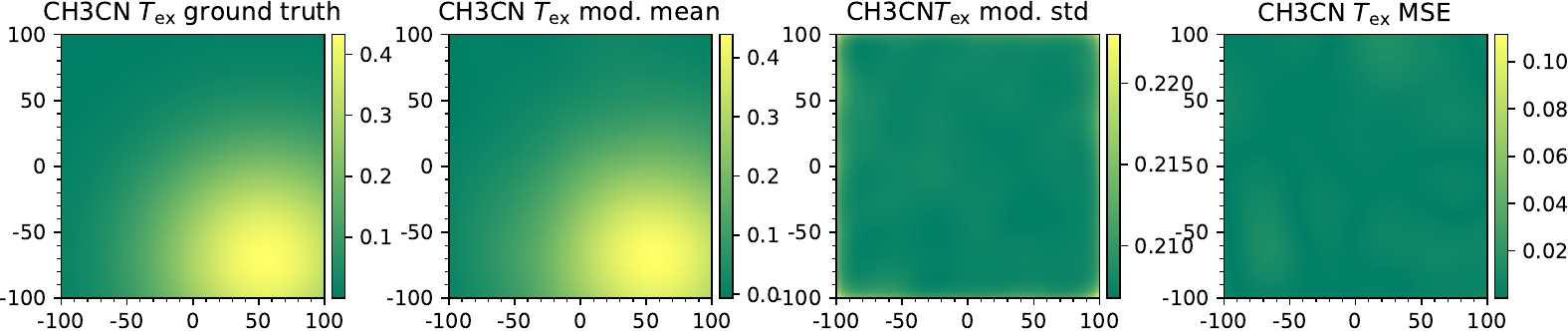}\\
 \hspace{-1.5cm}
  \includegraphics[scale=0.65]{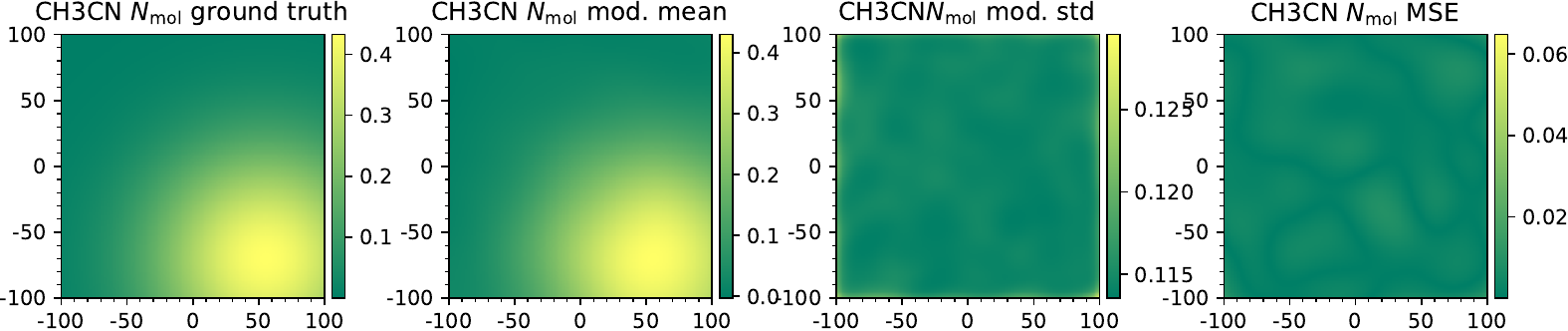}\\
   \includegraphics[scale=0.65]{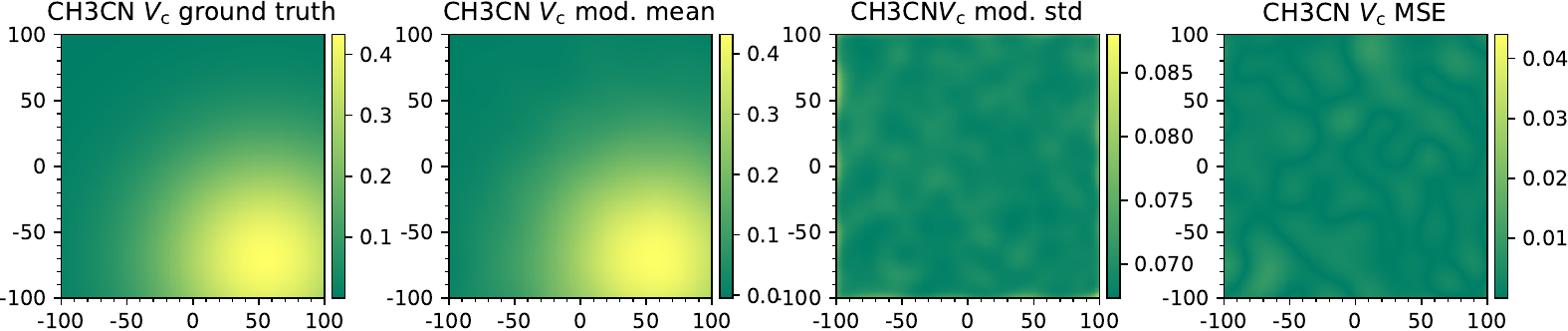}\\
   \includegraphics[scale=0.65]{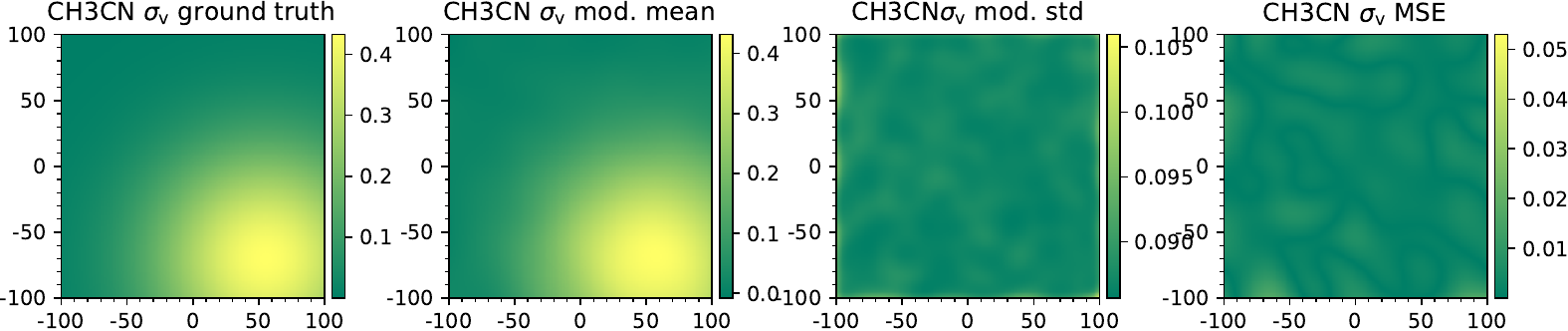}\\
  
\caption{Same as Fig. \ref{fig:compare_basq_ground_ch3oh}, but for CH$_{3}$CN parameters.}
    \label{fig:compare_basq_ground_ch3cn}
\end{figure*}

\begin{figure*}
 \includegraphics[scale=0.65]{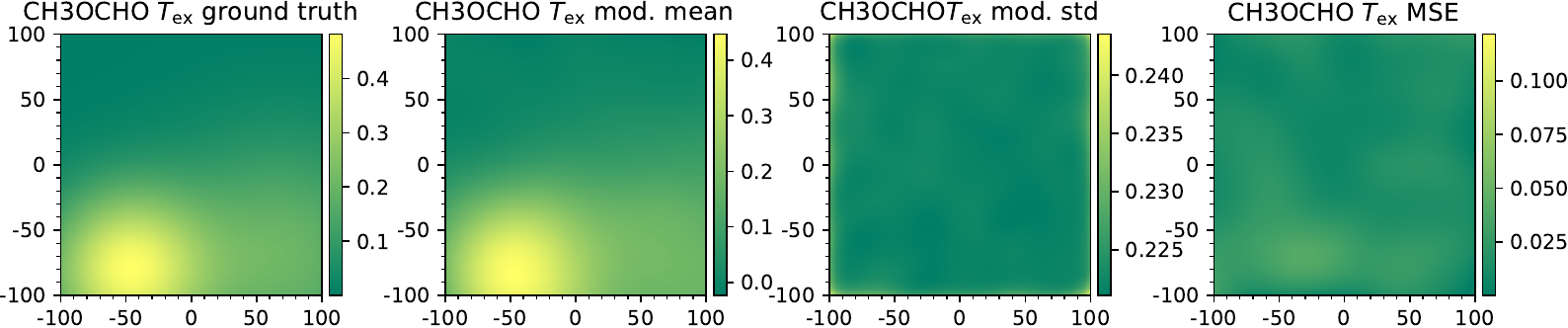}\\
 \hspace{-1.5cm}
  \includegraphics[scale=0.65]{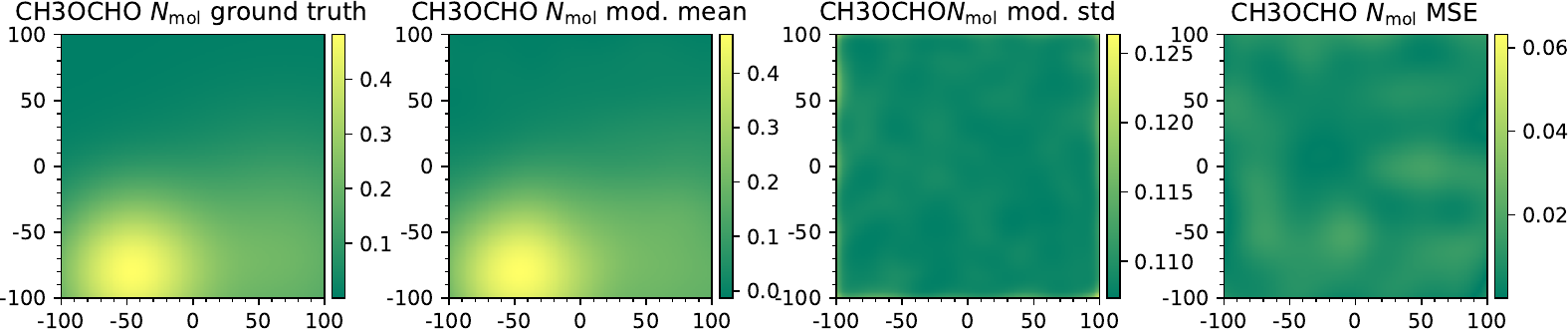}\\
   \includegraphics[scale=0.65]{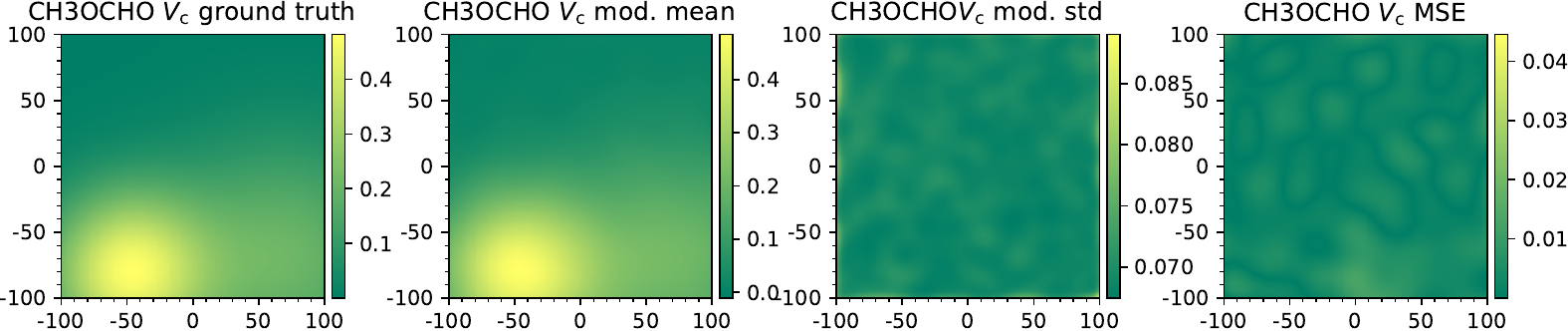}\\
   \includegraphics[scale=0.65]{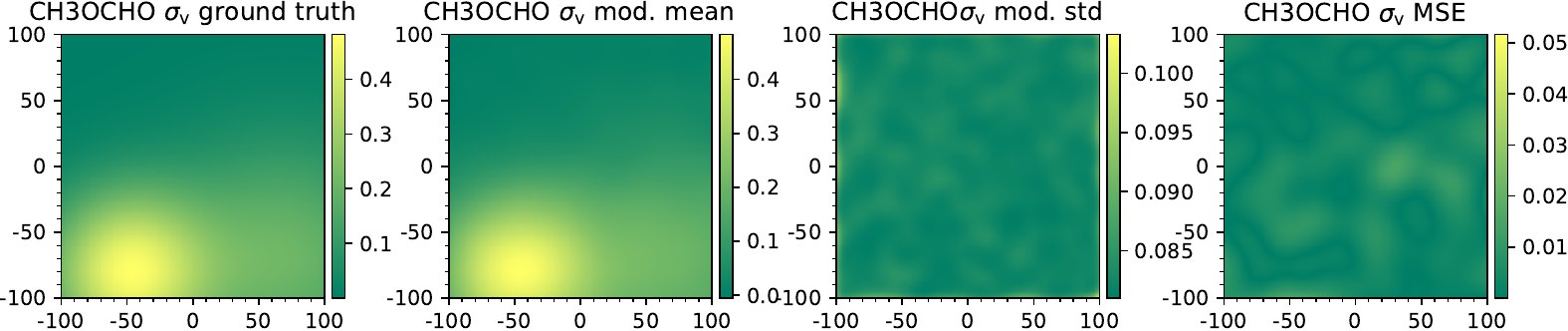}\\
  
\caption{Same as Fig. \ref{fig:compare_basq_ground_ch3oh}, but for CH$_{3}$OCHO parameters.}
    \label{fig:compare_basq_ground_ch3ocho}
\end{figure*}

\begin{landscape} % start landscape environment
\begin{table}[h]
\hspace{-1cm}
\begin{threeparttable}

\caption{List of molecules used for generating synthetic spectral cubes}
\begin{tabular}{cccccccccc}
\hline
2-atoms & 3-atoms & 4-atoms & 5-atoms & 6-atoms & 7-atoms & 8-atoms & 9-atoms &10-atoms\\
\hline
CO & HCN, v=0 (1) & H$_2$CNH & CH$3$Cl-35 &CH$_2$DOH & C$_2$H$_3$CN, v=0  &CH$_3$OCHO &CH$_3$OCH$_3$, v=0 &aGg'-(CH$_2$OH)$_2$\\
C-13-O (1)  & HNC, v=0 (1) & H$_2$CO & CH$_3$Cl-37 & HC(O)NH$_2$, v=0 &CH$_3$NCO, vb=0  &CH$_3$OC-13-HO, vt=0,1 & C$_2$H$_5$CN, v=0 &gGg'-(CH$_2$OH)$_2$\\
CO-18 (1)  & HC-13-N, v=0 (1) & HDCO & NH$_2$CN & HC(O)NH$_2$, v$_{12}$=1  &CH$_3$NCO, vb=1 &CH$_2$(OH)CHO &C$_2$H$_5$OH, v=0&CH$_3$CH$_2$CHO \\
CO-17  & HCN-15, v=0 (1) & H$_2$C-13-O &  c-C$_3$H$_2$ & HC-13-(O)NH$_2$  & CH$_3$CCH & CH$_2$(OH)C-13-HO &a-CH$_3$C-13-H$_2$OH &CH$_3$OCH$_2$OH \\
CS, v=0-4  & N$_2$O & H$_2$CO-17 & t-HCOOH  & HC(O)N-15-H$_2$  & CH$_3$CDO, vt=0,1&C-13-H$_2$(OH)CHO &a-C-13-H$_3$CH$_2$OH &CH$_3$COCH$_3$\\
CS-33, v=0,1 (2)  &HC-13-N-15, v=0 (1) & D$_2$CO & t-HC-13-OOH  & DC(O)NH$_2$  & CH$_2$DCHO, vt=0 & CH$_2$(OD)CHO &a-CH$_3$CH$2$OD&\\
CS-34, v=0,1 (2) &  DCO$^+$ (1) & H$_2$CO-18 &  H$_2$CC-13-O & CH$_3$SH, v=0-2 & CH$_3$C-13-HO,vt=0,1 & CHD(OH)CHO &a-CH$_3$CHDOH  \\
CS-36 (1) &H$_2$S (1)  & HNCO & H$_2$C-13-CO & cis-HC(O)NHD  & C-13-H$_3$CHO,vt=0,1 &CH$_2$(OH)CDO &a-a-CH$_2$DCH$_2$OH \\
SiO, v=0-10     &   HDS   &   H$_2$CS  & HDC$_2$O & trans-HC(O)NHD   & CH$_3$CHO &CH$_3$COOH, vt=0&a-s-CH$_2$DCH$2$OH          & \\
SO, v=0        &  HDS-34    &   HDCS  & HC$_3$N, v=0  &C-13-H$_3$CN, v=0    &   C$_2$H$_4$O          &HCOCH$_2$OH& CH$_3$CH=CH$_2$&\\
& OCS, v=0 (2) &D$_2$C-13-O & H$_2$CCO & C-13-H$_3$CN, v=0  &  & C$_2$H$_3$CHO
& \\
&  OCS, v$_2$=1  &DNCO &  H$_2$NC-13-N& CH$_3$C-13-N, v=0 &  &  \\
& OC-13-S (2)&HNC-13-O & HDNCN & CH$_3$CN-15, v=0  & &  \\
&  OCS-33 &HONO &DCOOH  & CH$_2$DCN &  &  \\
& OCS-34 &  & HCOOD &   C-13-H$_3$CN, v$_8$=1        &  &  \\
& O-18-CS& &HCOOD  &          CH$_3$C-13-N, v$_8$=1  & &  \\
& SO$_2$, v=0 &  &  &   CH$_3$CN-13, v$_8$=1  &  &  \\
& S-34-O$_2$ & &  &  CHD$_2$CN&  &  \\
& CCH, v=0 & & &  CH$_3$CN&  &  \\
&  & & &  CH$_3$CN, v$_8$=1&  &  \\
&  & & &CH$_3$NC & &  &  \\
&  & & &CH$_3$O-18-H, v=0-2 & &  &  \\
&  & & &C-13-H$_3$OH, v$_{\mathrm{t}}$=0,1 & &  &  \\
&  & & &CH$_3$OH, v$_{\mathrm{t}}$=0,1& &  &  \\
\hline
\end{tabular}
\label{tab:mollist}
\begin{tablenotes}
\small
\item[a] Numbers in parenthesis indicate the number of transitions within the frequency coverage, shown only for molecules with less than three transitions.
\end{tablenotes}

\end{threeparttable}
\end{table}
\end{landscape}

\end{appendix}

\end{document}